\documentclass[prd,showpacs,preprintnumbers,amsmath,amssymb]{revtex4}

\begin{document}

%\preprint{hep-ph/0301xxx}
%\preprint{Alberta Thy24-02}

\title{Nonequilibrium Evolution of Correlation Functions : A 
Canonical Approach}

\author{Supratim Sengupta}\email{sengupta@phys.ualberta.ca}
\author{F. C. Khanna}\email{khanna@phys.ualberta.ca}

\affiliation{Theoretical Physics Institute, Department of Physics,
University of Alberta, Edmonton, Alberta, Canada T6G 2J1}
\affiliation{TRIUMF, 4004 Wesbrook Mall, Vancouver, British
Columbia, Canada, V6T 2A3}

\author{Sang Pyo Kim}\email{sangkim@kunsan.ac.kr}

\affiliation{Department of Physics, Kunsan National University,
Kunsan 573-701, Korea}
\affiliation{Asia Pacific Center for
Theoretical Physics, Pohang 790-784, Korea}

\date{\today}

\begin{abstract}
We study nonequilibrium evolution in a self-interacting quantum
field theory invariant under space translation only by using a
canonical approach based on the recently developed Liouville-von
Neumann formalism. The method is first used to obtain the
correlation functions both in and beyond the Hartree approximation,
for the quantum mechanical analog of the $\phi^{4}$ model. The
technique involves representing the Hamiltonian in a Fock basis of
annihilation and creation operators. By separating it into a
solvable Gaussian part involving quadratic terms and a perturbation of
quartic terms, it is possible to find the improved vacuum state to 
any desired order. The correlation functions for the field theory are then
investigated in the Hartree approximation and those beyond the
Hartree approximation are obtained by finding the improved vacuum
state corrected up to ${\cal O}(\lambda^2)$. These correlation 
functions take into account next-to-leading and next-to-next-to-leading 
order effects in the coupling constant. We also use the Heisenberg
formalism to obtain the time evolution equations for the
equal-time, connected correlation functions beyond the leading order.
These equations are derived by including the connected 4-point
functions in the hierarchy. The resulting coupled set of equations
form a part of infinite hierarchy of coupled equations relating
the various connected n-point functions. The connection with other
approaches based on the path integral formalism is established and
the physical implications of the set of equations are discussed
with particular emphasis on thermalization.
\end{abstract}

\maketitle

\section{Introduction}

In the past few years, a lot of attention has been focused on the
investigation of classical and quantum fields evolving
out-of-equilibrium. Such interest is physically well motivated
because the very early history of the universe provides many
scenarios where nonequilibrium effects may have played an
important role. The reheating of the universe after inflation
\cite{linde}, formation and growth of domains in any generic
spontaneous symmetry breaking phase transition \cite{vega}, the
formation of topological defects
\cite{kbl,zurek,rivers,stephens,ssg,tanmay} and possible formation
of Quark-Gluon Plasma during the deconfinement transition or
Disoriented Chiral Condensates during the chiral phase transition
\cite{wilczek} are just some instances where the proper
understanding of the physical process may crucially depend on our
understanding of nonequilibrium quantum fields. The experimental
accessibility of some of these phenomena, like formation of quark
gluon plasma, made possible through heavy-ion colliders at RHIC
and LHC, has also been a strong motivating factor behind the
revival of interest in nonequilibrium evolution of quantum fields.
The development of new theoretical techniques and the availability
of more efficient computational resources have also made it
possible to investigate in some detail the nonlinear effects which
play such a crucial role in the evolution dynamics.

Another important issue in this context is that of thermalization
in closed quantum systems. Is it possible for macroscopic
irreversible behavior to manifest itself starting from microscopic
reversible (unitary) quantum dynamics
\cite{gleiser,boyan,greiner,berges}? In other words, would it be
possible for a closed quantum system to thermalize when it is
perturbed from its initial thermal state? Can the process of
thermalization be adequately described within the mean field
description? If not, what is the role of interactions
(non-linearities) in bringing about thermalization? Is it possible
to develop a consistent theoretical framework to address these
important issues? Some of these questions have only recently begun
to be addressed using newly developed theoretical tools for
dealing with nonequilibrium quantum fields
\cite{gleiser,boyan,greiner,berges,wett,cooper}.

Till recently, the issue of thermalization of a closed quantum
system was typically addressed by a separation of the system into
a sub-system made up of the non-thermal soft modes (with longer
thermalization time-scales) and an environment consisting of the
thermal hard modes (which thermalize on much shorter time scales
compared to the soft modes) \cite{gleiser,boyan,greiner}. The
subsequent interaction between the soft modes and the environment
(often treated stochastically) leads to the eventual thermalization
of the sub-system  made up of soft modes. Mean field theory
(Hartree approximation) has also been extensively and successfully
used to study the dynamics of nonequilibrium quantum fields
\cite{jackiw1,habib,boyan2,baacke} and has yielded valuable
insights into the early-time dynamical behavior. However, since
the mean field (Hartree) approximation is essentially a linear
approximation, it fails to correctly capture long-time dynamical
behavior where nonlinear effects play a dominant role. Moreover,
in the mean field approximation, the different field modes
interact with a spatially homogeneous mean field with equal
strength and therefore the effects of direct scattering (which is
responsible for redistribution and eventual equipartition of
energy among different modes) are neglected. Hence thermalization
of the system cannot be achieved \cite{berges,cooper}. Recently,
there have been some attempts \cite{smit,betten} to explore the
possibility of thermalization within the mean field scheme by
considering spatially inhomogeneous mean fields. The motivation
was to see if the scattering of the fluctuation modes with the
modes of the inhomogeneous mean field aids in thermalization.
Salle, Smit and Vink \cite{smit} proposed a density matrix
describing an ensemble of pure Gaussian initial states. By
defining the mean field to be an ensemble average over a set of
Gaussian density matrices, and considering the interaction between
the quantum fluctuation modes and the inhomogeneous mean field,
they showed approximate thermalization is observed over
intermediate time scales in the sense of particle distribution of
low momentum modes approaching a Bose-Einstein form for a
spontaneously broken theory. For long times, the particle
distributions were found to tend towards a classical Boltzmann
form. However, in a contrasting study, Bettencourt et.al.
\cite{betten} showed that by considering the dynamics in the
presence of a spatially inhomogeneous mean field, but {\it
without} ensemble averaging over a certain class of initial
conditions, they found that the Hartree approximation fails to
establish a thermal Bose-Einstein particle distribution even at
late times. In a more recent paper \cite{salle} Salle and Smit
pointed out that the energy densities used in the simulations of
\cite{betten} were not large enough to ensure even approximate
thermalization in the BE sense over the time scales observed. They
found that by choosing a large enough energy density and a
different set of initial conditions, BE behaviour was exhibited by
the low momentum modes at intermediate time scales, even for a
symmetric theory. However, even the approximate thermalization
time scale was much larger in comparison to that of a
spontaneously broken theory. In none of these works, quantum
thermalization was observed for all modes and all energy
densities, over the time scales observed.

In order to understand the complicated process of thermalization
of closed quantum field systems, one needs to go beyond the mean
field (Hartree or leading order Large - N) approximation schemes.
This has been attempted recently by many different groups
\cite{berges,wett,cooper,yaffe,aarts}. More recently,
time-reversal invariant equations for correlation functions of
nonequilibrium scalar fields have been derived \cite{berges} based
on a 3-loop expansion of the 2PI effective action
\cite{jackiw,hu}. This method has also been extended to O(N)
symmetric field theories by carrying out a systematic 1/N
expansion of the 2PI effective action. The 3-loop approximation
(for the single real scalar field case) and the
next-to-leading-order (NLO) expansion for the O(N) model
incorporates the effect of direct scattering and shows promising
evidence of late-time thermalization. A detailed analysis of the
nonequilibrium dynamics shows the evidence of three distinct
temporal regimes \cite{berges} characterized by an early-time
exponential damping, an intermediate hydrodynamic regime and a
late-time regime manifest through exponential approach to thermal
equilibrium. An alternative approximation scheme has been
developed by Cooper and collaborators focusing on resummation
methods based on the exact Schwinger-Dyson equations
\cite{cooper}. This scheme called the bare vertex approximation
scheme (BVA) (first studied by Kraichnan \cite{kra}) involves
obtaining the exact Schwinger-Dyson equations and then neglecting
the vertex corrections. All the methods discussed above are based
on path-integral techniques within the Keldysh-Schwinger closed
time path formalism \cite{schwinger} and the basic dynamical
equations can be quite complicated to derive.

Our main aim in this paper is to make use of canonical formalism
to study the nonequilibrium evolution of quantum fields. 
%and establish a connection with other approaches, most of which are
%based on the functional integral formalism. 
The Liouville-von Neumann (LvN) formalism was first applied 
(within the Hartree approximation scheme) 
to study the early-time growth of domains during a quenched second order
phase transition \cite{kim-lee2}. The LvN formalism which solves
directly the quantum LvN equation, is another quantum picture
\cite{lewis,kim} besides the Schr\"{o}dinger and Heisenberg
pictures. Further, it is shown that the LvN formalism provides a
convenient and powerful method for analysing nonequilibrium systems such 
as time-dependent oscillators \cite{lewis,kim,kim-kim} and quenched
phase transitions \cite{kim-lee2,ksk,kim-lee,ramos}. We use this
formalism to obtain the Gaussian vacuum and thermal evolution
equations in the Hartree approximation. We develop a method for 
obtaining the improved vaccum state and explicitly show how to 
obtain equations for the first and second order coefficients which
lead to the improved vacuum state correct to ${\cal O}(\lambda^2)$.  
We then derive expressions for the 2-point functions by taking the expectation 
value of products of field operators with respect to this improved vacuum state.This method clearly indicates that non-Gaussian effects first make their 
appearance at ${\cal O}(\lambda^2)$.

An alternative canonical approach based on the Heisenberg formalism is 
then employed to obtain a set of nonequilibrium evolution equations for
the correlation functions. After clarifying the relation between the LvN 
formalism and the Heisenberg-picture, we take the vacuum expectation values 
of the Heisenberg equations for all possible combinations of products of 
field operators and obtain a hierarchy of coupled equations for the ordinary 
n-point correlators. To obtain the dynamical equations for the connected 
equal-time correlators beyond the leading order, we make use of the method of 
cluster expansion which allows us to express the ordinary n-point correlators in 
terms of their connected counterparts. This method provides an alternative
non-perturbative approach for going beyond the mean field approximation.
We compare our results with those obtained in the literature using
other methods \cite{cw1,yaffe}. In this context we consider two
recent approaches used to obtain the evolution equations for the
partition function \cite{cw1} and correlators in a quantum
mechanical model discussed by Ryzhov and Yaffe (RY)\cite{yaffe}.
We show that the canonical approach used in this paper yields the
appropriate evolution equations, and thereby establish a connection
between various approaches for obtaining the nonequilibrium evolution
equations for {\it equal-time} correlation functions.

The paper is organized as follows. In the next section we briefly outline
the  Liouville-von Neumann (LvN) formalism which will be subsequently used
to study the nonequilibrium dynamics of the correlation functions in and 
beyond the Hartree approximation. In Sec. III, we describe in detail, a
quantum-mechanical model of the anharmonic oscillators and use
the LvN formalism to obtain the evolution equations for the vacuum and
thermal correlation functions in the Hartree approximation. In Sec. IV, 
we develop the LvN formalism to study the non-Gaussian dynamics of the 
quantume mechanical anharmonic oscillator model. This section sets the 
stage for the application of the LvN formalism to study non-Gaussian 
dynamics in the more complicated field theory model. The issue of stability 
of the LvN method, in the context of the anharmonic oscillator, is 
discussed in Sec.IV.B. In Sec. V, the evolution equations for a 
{\it nonlinear} self-interacting scalar field theory are first derived 
in the Hartree approximation, both for the nonequilibrium as well as 
for the thermal equilibrium case. Nonequilibrium evolution beyond leading 
order is discussed in detail Sec. VI which contains 
the most important results of this paper. The LvN formalism is used to 
investigate nonequilibrium dynamics beyond the leading order in Sec. VI.A. 
In Sec. VI.B. we make use of the Heisenberg formalism to obtain a hierarchy
of nonequilibrium evolution equations for the connected n-point functions.    
A comparison between our approach and other methods used in the literature
is carried out in Sec. VII. Sec. VIII contains a summary of our main
results and a discussion of the physical implications of the
results in the context of issues like thermalization in quantum
field theory and phase transitions. The method of cluster
expansion \cite{peter} is outlined in Appendix A, where we also
write down the set of equations for the connected correlators in
configuration space. The equivalence between the RY method and the
LvN formalism is established in Appendix B.

\section{Liouville-von Neumann Picture for Time-Dependent Systems}

The systems under study in this paper have Hamiltonians whose
coupling constants (parameters) depend on time explicitly. These
systems describe nonequilibrium processes in the sense that the
Hamiltonians do not give the correct density operators $e^{- \beta
\hat{H} (t)}/Z_{\rm H}$. To properly find the Hilbert space and
density operators for such time-dependent systems, we have to
clarify the picture of the quantum evolution. Let the Hamiltonian
be defined in terms of the Schr\"{o}dinger operators for a quantum
mechanical system
\begin{equation}
\hat{H} (t) = \hat{H} (\hat{p}_{\rm S}, \hat{q}_{\rm S}, t), \label{ham1}
\end{equation}
and for a quantum field theory
\begin{equation}
\hat{H} (t) = \hat{H} (\hat{\pi}_{\rm S}, \hat{\phi}_{\rm S}, t).\label{ham2}
\end{equation}
Here the systems depend on time only through time-dependent
coupling constants (parameters). In addition to the well-known
Schr\"{o}dinger, Heisenberg and interaction pictures there is
another picture for such nonequilibrium systems \cite{lewis,kim}.

Firstly, in the Schr\"{o}dinger picture, the (functional)
Schr\"{o}dinger equation (in unit of $\hbar = 1$)
\begin{equation}
i  \frac{\partial}{\partial t} \vert \Psi (t) \rangle =
\hat{H} (t) \vert \Psi(t)\rangle  \label{sch eq}
\end{equation}
has the exact quantum state
\begin{equation}
\vert \Psi (t) \rangle = \hat{U}(t) \vert \Psi \rangle_{\rm S}
\end{equation}
determined by the unitary evolution operator
\begin{equation}
i  \frac{\partial}{\partial t} \hat{U} (t) = \hat{H} (t)
\hat{U} (t).
\end{equation}
For the time-dependent case in contrast with the time-independent
case, it is not easy to find the evolution operator, which is
formally defined as
\begin{equation}
\hat{U} (t) = {\rm T} \exp \Bigl[- i \int^t \hat{H} (t') dt'
\Bigr], \label{ev op}
\end{equation}
where T denotes the standard time-ordered operator. Secondly, the
Heisenberg operators
\begin{equation}
\hat{O}_{\rm H} (t) = \hat{U}^{\dagger} (t) \hat{O}_{\rm S} \hat{U} (t)
\label{heis op}
\end{equation}
satisfy the Heisenberg equation of motion
\begin{equation}
i \frac{\partial}{\partial t} \hat{O}_{\rm H} (t) + [\hat{H}_{\rm H} (t),
\hat{O}_{\rm H} (t) ] = 0, \label{heis eq}
\end{equation}
where $\hat{H}_{\rm H} (t)$ is the time-dependent Hamiltonian
Heisenberg operator. The Heisenberg operator $\hat{H}_{\rm H} (t)$
is simply given by replacing the Schr\"{o}dinger operators
$\hat{p}_{\rm S}$ and $\hat{q}_{\rm S}$ by $\hat{p}_{\rm H} (t)$
and $\hat{q}_{\rm H} (t)$ in Eq. (\ref{ham1}) and $\hat{\pi}_{\rm
S}$ and $\hat{\phi}_{\rm S}$ by $\hat{\pi}_{\rm H} (t)$ and
$\hat{\phi}_{\rm H} (t)$ in Eq. (\ref{ham2}). However, the
explicit form of
 $\hat{p}_{\rm H} (t)$, $\hat{q}_{\rm H} (t)$ and  $\hat{\pi}_{\rm H} (t)$,
$\hat{\phi}_{\rm H} (t)$ requires either the exact knowledge of
the evolution operator (\ref{ev op}) in advance or the solution to
Eq. (\ref{heis eq}). Thirdly, we may introduce the Liouville
operators
\begin{equation}
\hat{O}_{\rm L} (t) = \hat{U} (t) \hat{O}_{\rm S} \hat{U}^{\dagger} (t).
\label{liou op}
\end{equation}
Thus the Liouville operators evolve the Schr\"{o}dinger operators
backward in time \cite{balian}. It follows then that the Liouville
operators satisfy the quantum LvN equation
\begin{equation}
i \frac{\partial}{\partial t} \hat{O}_{\rm L} (t) + [\hat{O}_{\rm L}
(t), \hat{H} (t)] = 0. \label{ln eq1}
\end{equation}
We can show that any eigenstate of the Liouville operator
\begin{equation}
\hat{O}_{\rm L} (t) \vert \Psi_{\lambda}, t \rangle = \lambda \vert
\Psi_{\lambda}, t \rangle,
\end{equation}
where $\lambda$ is the eigenvalue of the corresponding Schr\"{o}dinger
operator $\hat{O}_{\rm S}$, satisfies the Schr\"{o}dinger equation
(\ref{sch eq}).
In fact, it follows that \cite{lewis}
\begin{equation}
\vert \Psi(t) \rangle = \sum_{\lambda} C_{\lambda}
e^{ i \int{dt \langle
\Psi_{\lambda},t \vert (i \frac{\partial}{\partial t} -
\hat{H}(t))\vert \Psi_{\lambda},t \rangle} } \vert \Psi_{\lambda}, t \rangle
\label{ex st}
\end{equation}
where $C_{\lambda}$'s are constants.

The essential idea of the LvN method
\cite{kim-lee2,lewis,kim,kim-kim,ksk} is that the quantum LvN
equation provides all the quantum and statistical information of
nonequilibrium systems. Technically the linearity of the LvN
equation allows any functional of an operator satisfying the LvN
equation to be another operator. Thus we may use some suitable
operator $\hat{O}_{\rm L} (t)$ to define the density operator
$\hat{\rho} (t) = e^{- \beta\hat{O}_{\rm L} (t)}/Z_{O}$ for the
time-dependent system provided  $\hat{O}_{\rm L} (t)$ satisfies
the LvN equation. In this sense the LvN method unifies quantum
statistical mechanics with quantum mechanics. The LvN method
treats the time-dependent, nonequilibrium system exactly in the
same way as the time-independent, equilibrium one. Moreover, the
LvN method can be applied to nonequilibrium fermion systems with a
minimal modification \cite{kim-khanna}.

\section{Anharmonic Oscillator in the Hartree Approximation}

As a precursor to the investigation of $\phi^{4}$-field theory, we
apply the LvN method to a simple quantum mechanical model of
anharmonic oscillators
\begin{equation}
\hat{H} = \frac{1}{2} \hat{p}^2  \pm \frac{\omega^2}{2}\hat{x}^2 +
\frac{\lambda}{4!}\hat{x}^4, \label{anh osc}
\end{equation}
and derive the evolution equations for the coherent state
expectation value of position and momentum variables as well as
the subtracted 2-point correlators. The anharmonic oscillator with
the lower sign is a quantum mechanical analog for the second order
phase transition. All the time-dependent operators in Secs. III,
IV and V will denote Liouville operators, whose subscript L will
be dropped.

The main idea behind the LvN method is to require the pair of
invariant operators \cite{kim-kim,manko} defined as
\begin{equation}
\hat{a}(t) = i [u^{*}(t)\hat{p} -
\dot{u}^{*}(t)\hat{x}], \quad \hat{a}^{\dagger}(t) = -
i [u(t)\hat{p} - \dot{u}(t)\hat{x}], \label{eq
aa}
\end{equation}
to satisfy the LvN equation i.e.
\begin{equation}
i \frac{\partial \hat{a}}{\partial t} + [\hat{a},\hat{H}(t)]
= 0, \quad i \frac{\partial \hat{a}^{\dagger}}{\partial t} +
[\hat{a}^{\dagger},\hat{H}(t)] = 0. \label{eq aad}
\end{equation}
Here $u(t)$ and $u^{*}(t)$ are auxiliary variables in terms of
which the 2-point correlators will be expressed. The invariant
operators may be made the annihilation and creation operators
satisfying the standard commutation relation at equal times
\begin{equation}
[ \hat{a} (t), \hat{a}^{\dagger} (t) ] = 1,
\end{equation}
which leads to the Wronskian condition
\begin{equation}
\dot{u}^* (t) u(t) - u^* (t) \dot{u} (t) = i. \label{wron con}
\end{equation}
All the other commutation relation vanish
\begin{equation}
[ \hat{a} (t), \hat{a} (t) ] = [ \hat{a}^{\dagger} (t),
\hat{a}^{\dagger} (t) ] = 0.
\end{equation}
Equation (\ref{eq aa}) can be inverted to express the position and
momentum operators $\hat{x}$ and $\hat{p}$ in terms of the
annihilation and creation operators as
\begin{eqnarray}
\hat{x}(t) &=& u(t)\hat{a}(t) +
u^{*}(t)\hat{a}^{\dagger}(t), \nonumber \\ \hat{p}(t) &=&
\dot{u}(t)\hat{a}(t) +
\dot{u}^{*}(t)\hat{a}^{\dagger}(t). \label{eq bb}
\end{eqnarray}
The coherent state is defined either as the eigenstate of $\hat{a}
(t)$
\begin{equation}
\hat{a} (t) \vert \alpha, t \rangle = \alpha \vert \alpha, t
\rangle, \label{coh st}
\end{equation}
with a complex eigenvalue $\alpha$ or as the displaced state of
the vacuum state given by
\begin{equation}
\vert \alpha, t \rangle = \hat{D}^{\dagger} (\alpha) \vert 0, t
\rangle = e^{ - \alpha^* \alpha/2} \sum_{n = 0}^{\infty}
\frac{\alpha^n}{\sqrt{n!}} \vert n, t \rangle,
\end{equation}
where $\hat{D}$ is the displacement operator
\begin{equation}
\hat{D} (\alpha) = e^{ - \alpha \hat{a}^{\dagger} (t) + \alpha^*
\hat{a} (t)}. \label{dis op}
\end{equation}
The coherent state can also be found using the variational
principle \cite{rajagopal2}.

The coherent state then leads to the expectation value of
$\hat{x}$, $\hat{p}$, $\hat{x}^2$ and $\hat{p}^2$:
\begin{eqnarray}
\bar{x} \equiv \langle \alpha,t \vert \hat{x} \vert \alpha,t
\rangle &=& \alpha u(t) + \alpha^{*}u^{*}(t),
\nonumber
\\ \bar{p} \equiv \langle \alpha,t \vert \hat{p} \vert \alpha,t
\rangle &=& \alpha \dot{u}(t) +
\alpha^{*}\dot{u}^{*}(t), \nonumber \\ \langle \alpha,t \vert
\hat{x}^2 \vert \alpha,t \rangle &=& \bar{x}^2 +  u^{*}(t) u
(t), \nonumber \\ \langle\alpha,t \vert \hat{p}^2 \vert
\alpha,t\rangle &=& \bar{p}^2 + \dot{u}^{*} (t) \dot{u}(t).
\label{eq cc}
\end{eqnarray}
The subtracted 2-point correlators \cite{yaffe} defined below are
then given by
\begin{eqnarray}
g_{xx}(t) &=& \langle \hat{x}^2 \rangle  - \bar{x}^2 = 
u^{*}(t) u(t), \nonumber \\ g_{pp}(t) &=& \langle \hat{p}^2
\rangle - \bar{p}^2 =  \dot{u}^{*}(t) \dot{u}(t), \nonumber
\\ g_{xp}(t) &=& \langle \hat{x}\hat{p} \rangle  - \bar{x}\bar{p}
=  \dot{u}^{*}(t)u(t), \nonumber \\ g_{px} (t) &=&
g_{xp}^{*}(t)  =  u^{*}(t) \dot{u}(t), \label{eq cr}
\end{eqnarray}
from which we obtain the evolution equations for the 2-point
correlators
\begin{eqnarray}
\dot{g}_{xx} (t) &=& g_{xp} (t) + g_{px} (t), \nonumber \\
\dot{g}_{pp} (t) &=& \ddot{u}^{*} (t) \dot{u} (t) +
\dot{u}^{*} (t) \ddot{u}(t), \nonumber \\ \dot{g}_{xp} (t) &=&
\ddot{u}^{*} (t) u(t) + \dot{u}^{*}(t) \dot{u}(t).
\label{eq evo}
\end{eqnarray}

\subsection{Correlation Functions in Coherent State}

The expectation value of the Hamiltonian with respect to the
coherent state (\ref{coh st}) leads to an effective Hamiltonian
\begin{equation}
H_{\rm eff} \equiv \langle \hat{H} \rangle_{\rm cs} =
\frac{1}{2}\bar{p}^2 \pm \frac{\omega^2}{2}\bar{x}^2 +
\frac{\lambda}{4!}\bar{x}^4 + \frac{1}{2}g^{(2)}_{xx}(\pm \omega^2
+ \frac{\lambda}{2}\bar{x}^2) + \frac{3\lambda}{4!}g^{(2)2}_{xx} +
{\cal O} (\lambda^3 ).
\end{equation}
The auxiliary field $u(t)$, after differentiating Eq. (\ref{eq
aad}) with respect to $\hat{x}$ and taking the coherent state
expectation value, satisfies the equation
\begin{equation}
\ddot{u}(t) \pm \omega^2 u(t) + \langle \alpha,t \vert
\frac{\delta^2 \hat{H}}{\delta \hat{x}^2}\vert \alpha,t \rangle
u(t) = 0, \label{gen eq}
\end{equation}
so does its complex conjugate $u^*(t)$. Inserting the expectation
values obtained in Eq. (\ref{eq cc}) into Eq. (\ref{gen eq}), one
obtains the equation for the complex $u(t)$
\begin{equation}
\ddot{u}(t) + \Bigl[\pm \omega^2 + \frac{\lambda}{2}\bar{x}^2 +
\frac{\lambda}{2} u^{*}u \Bigr] u(t) = 0. \label{eq av}
\end{equation}
Using the above equation and the definition, Eq. (\ref{eq cr}), of
the subtracted 2-point correlators, we get the following coupled
set of evolution equations for the subtracted 2-point correlators
\begin{eqnarray}
\dot{g}_{xx}(t) &=& g_{xp}(t) + g_{px}(t), \nonumber \\
\dot{g}_{pp}(t) &=& - \Bigl(\pm \omega^2 +
\frac{\lambda}{2}\bar{x}^2 \Bigr )(g_{xp} + g_{px}) + {\cal O}
(\lambda^2),  \nonumber\\ \dot{g}_{xp}(t) &=& g_{pp} -
g_{xx} \Bigl(\pm \omega^2 + \frac{\lambda}{2}\bar{x}^2 \Bigr) +
{\cal O} ( \lambda^2). \label{eq aho}
\end{eqnarray}
It is important to note that, these equations are correct only up
to ${\cal O}(\lambda^2)$ except for the first one which is
exact. This is due to the quartic term appearing in the potential,
unlike the case of the simple harmonic oscillator where the
quadratic (Gaussian) form of the Hamiltonian leads to the
vanishing of all correlators greater than second order. In this
case, even if one starts from a Gaussian state peaked at $\bar{x}
= 0$ for which all the correlators greater than second order
vanished, the subsequent evolution of the coupled set of equations
would induce the appearance of non-vanishing values for higher
order correlators at ${\cal O}(\lambda^3)$ and higher
\cite{yaffe}. This will be discussed later when we obtain the
evolution equations for a $\phi^4$-field theory beyond the leading
order.

\subsection{Correlation Functions in Thermal State}

For the case of the anharmonic oscillator in an initial thermal
equilibrium with the positive sign for the unbroken symmetry, the
evolution equations for the auxiliary variable and the 2-point
correlators up to ${\cal O}(\lambda^2)$ can be obtained by
taking the expectation value with respect to the coherent-thermal
state
\begin{equation}
\hat{\rho}_{\rm CT} = \frac{1}{Z_{\rm CT}} \exp[ - \beta \{
\Omega \hat{a}^{\dagger} (t) \hat{a} (t) + \delta
\hat{a}^{\dagger} (t) + \delta^* \hat{a}(t) + \epsilon_0\}],
\label{den op}
\end{equation}
where $\epsilon_0 = \Omega/2 + |\delta|^2/(\Omega)$.
As $\hat{a}(t)$ and $\hat{a}^{\dagger} (t)$ satisfy approximately
the quantum LvN equation, Eq.(\ref{eq aad}), so does the density
operator (\ref{den op}). We choose $\Omega$ to satisfy the gap
equation for the unbroken symmetry
\begin{equation}
\Omega^2 = \omega^2 + \frac{\lambda}{4 \Omega}.
\end{equation}
As discussed in \cite{kim-lee2}, the displacement operator
(\ref{dis op}) with $\alpha = \delta /(\Omega)$ unitarily
transforms the coherent-thermal density matrix to a thermal one
\begin{equation}
\hat{D}^{\dagger} (\alpha) \hat{\rho}_{\rm CT}\hat{D}(\alpha) =
\frac{1}{Z_{\rm T}} \exp[ - \beta \Omega \hat{a}^{\dagger}
(t) \hat{a} (t)] = \hat{\rho}_{\rm T}.
\end{equation}
By making use of the unitary transformation
\begin{eqnarray}
\hat{D}^{\dagger}(\alpha)\hat{a} (t) \hat{D}(\alpha) = \hat{a} (t)
- \alpha, \quad \hat{D}^{\dagger}(\alpha)\hat{a}^{\dagger} (t)
\hat{D}(\alpha) = \hat{a}^{\dagger} (t) - \alpha^*,
\end{eqnarray}
it is easy to show that
\begin{eqnarray}
\langle \hat{x}^2\rangle_{\rm CT} = \bar{x}^2 + 
\coth\Bigl(\frac{\beta \Omega}{2}\Bigl) u^{*}(t)u(t),
\nonumber
\\ \langle\hat{p}^2\rangle_{\rm CT} = \bar{p}^2 +
\coth\Bigl(\frac{\beta \Omega}{2}\Bigl)
\dot{u}^{*}(t)\dot{u}(t),
\end{eqnarray}
where now
\begin{eqnarray}
\bar{x} = \langle \alpha, t \vert \hat{x} \vert \alpha, t \rangle
= - (\alpha u + \alpha^* u^*), \nonumber\\ \bar{p} =
\langle \alpha, t \vert \hat{p} \vert \alpha, t \rangle = -
 (\alpha \dot{u} + \alpha^* \dot{u}^*).
\end{eqnarray}
The expectation value of Eq. (\ref{eq aad}) with respect to the
coherent-thermal state (\ref{den op}) leads to the equation for
the complex $u(t)$:
\begin{equation}
\ddot{u}(t) + \Bigl[\pm \omega^2 + \frac{\lambda}{2}\bar{x}^2 +
\frac{\lambda}{2}
\coth\Bigl(\frac{\beta \Omega}{2}\Bigl) u^{*}u \Bigr ]u(t) =
0. \label{eq thermu}
\end{equation}
The corresponding thermal 2-point correlators are then given by
\begin{eqnarray}
g_{{\rm T} xx}(t) &=& \langle \hat{x}^2\rangle_{\rm CT} -
\bar{x}^2 = \coth\Bigl(\frac{\beta \Omega}{2}\Bigl)
u^{*}(t)u(t), \nonumber
\\ g_{{\rm T} pp}(t) &=& \langle\hat{p}^2\rangle_{\rm CT} -
\bar{p}^2 = \coth\Bigl(\frac{\beta \Omega}{2}\Bigl)
\dot{u}^{*}(t)\dot{u}(t), \nonumber
\\ g_{{\rm T} xp} (t) &=& \langle\hat{x}\hat{p}\rangle_{\rm CT} -
\bar{x}\bar{p} = \coth\Bigl(\frac{\beta \Omega}{2}\Bigl)
\dot{u}^{*}(t)u(t), \nonumber
\\ g_{{\rm T} px} (t) &=& g_{{\rm T} xp}^{*}(t) = 
\coth\Bigl(\frac{\beta \Omega}{2}\Bigl) u^{*}(t) \dot{u}(t),
\label{eq ctr}
\end{eqnarray}
from which we obtain the evolution equations for the thermal
2-point correlators
\begin{eqnarray}
\dot{g}_{{\rm T} xx}(t) &=& g_{{\rm T} xp}(t) + g_{{\rm T} px}
(t), \nonumber \\ \dot{g}_{{\rm T} pp}(t) &=& - \Bigl[\pm \omega^2
+ \frac{\lambda}{2}\bar{x}^2 +
\frac{\lambda}{2}\coth\Bigl(\frac{\beta \Omega}{2}\Bigl)
u^{*}u \Bigl] (g_{{\rm T} xp} + g_{{\rm T} px}) + {\cal
O}(\lambda^2), \nonumber\\ \dot{g}_{{\rm T} xp}(t) &=&
g_{{\rm T} pp} - g_{{\rm T} xx} \Bigl[\pm \omega^2 +
\frac{\lambda}{2}\bar{x}^2 + \frac{\lambda}{2}
\coth\Bigl(\frac{\beta \Omega}{2}\Bigl)u^{*}u \Bigl] + {\cal
O}(\lambda^2). \label{eq taho}
\end{eqnarray}
In the $T=0$ limit, Eqs. (\ref{eq ctr}) and (\ref{eq taho}) reduce
to the evolution equations (\ref{eq av}) and (\ref{eq aho}).
Expressions for the subtracted 2-point correlators can then be
obtained by solving Eqs. (\ref{eq thermu}) for $u(t)$
perturbatively in powers of $\lambda$.

\section{Anharmonic Oscillator beyond the Hartree Approximation}

In this section we apply the LvN method to the anharmonic
oscillator, in Sec. III, to go beyond the Hartree approximation.
Our stratagem is to represent the full Hamiltonian (\ref{anh osc})
in the Fock space basis (\ref{eq aa}) and to follow the standard
perturbation theory by taking the quadratic terms as an
unperturbed part and the quartic terms as a perturbation. For the
convenience of computation, we express the representation in the
normal ordering where all $\hat{a}^{\dagger}$ stand to the left of
$\hat{a}$.

We divide $\hat{H}$ into a Gaussian part $\hat{H}_{\rm G}$, the
quadratic part, and a perturbation Hamiltonian $\hat{H}_{\rm P}$,
the remaining quartic part,
\begin{eqnarray}
\hat{H} = \hat{H}_{\rm G} + \lambda \hat{H}_{\rm P},
\end{eqnarray}
where
\begin{eqnarray}
 \hat{H}_{\rm G} &=& \frac{1}{2} :\hat{p}^2: \pm \frac{\omega^2}{2}
 :\hat{x}^2: + \frac{6 \lambda}{4!} \langle \hat{x}^2
\rangle_{\rm G} :\hat{x}^2:  + E_{\rm G}, \label{2-p}\\
\hat{H}_{\rm P} &=&  \frac{1}{4!} :\hat{x}^4:. \label{4-p}
\end{eqnarray}
Here $E_{\rm G}$ is the vacuum expectation value
\begin{equation}
E_{\rm G} = \frac{1}{2} \langle \hat{p}^2 \rangle_{\rm G} \pm
\frac{\omega^2}{2} \langle \hat{x}^2 \rangle_{\rm G} + \frac{3
\lambda}{4!} \langle \hat{x}^2 \rangle^2_{\rm G}.
\end{equation}
Using the normal ordered operators
\begin{equation}
: (u \hat{a} + u^* \hat{a}^{\dagger})^n :~ = \sum_{k = 0}^{n}
\frac{n!}{k! (n-k)!} ~u^{*(n-k)} u^k \hat{a}^{\dagger (n-k) }
\hat{a}^k,
\end{equation}
we obtain the expectation value with respect to the Gaussian
vacuum state that is annihilated by $\hat{a} (t)$
\begin{equation}
E_{\rm  G} (t) = \frac{1}{2} \Bigl[ \dot{u}^* \dot{u} \pm
\omega^2 u^* u + \frac{\lambda}{4} ( u^* u )^2 \Bigr].
\label{vac en}
\end{equation}

A few comments are in order. First, the separation of the
Hamiltonian into the quadratic and quartic parts in Eqs.
(\ref{2-p}) and (\ref{4-p}) is reminiscent of the
Caswell-Killingbeck method \cite{caswell}, which separates the
Hamiltonian into a solvable part and a perturbation. In fact, as
we shall show below, the quadratic part  is  solvable via the LvN
method even for explicitly time-dependent systems. Second, the
quadratic part (\ref{2-p}) involves a term proportional to the
coupling constant $\lambda$, which makes any perturbation
theory based on it reliable even in the strong coupling limit of
$\lambda$. This term is the same as the Hartree approximation,
$\hat{q}^4 \rightarrow 6 \langle \hat{q}^2 \rangle \hat{q}^2$. As
will be shown below, the wave function(al)s of the Hamiltonian
(\ref{2-p}) are the same as those from the Gaussian effective
potential method by Chang and Stevenson \cite{chang}. The vacuum
state is also the same as the Hartree approximation. The
equivalence of the vacuum state between the LvN and Gaussian
effective potential will be shown below (see also Ref.
\cite{kim-lee2}).

The Hartree approximation in Sec. III is equivalent to simply
using the truncated quadratic part, $\hat{H}_{\rm G}$, as the
unperturbed part. Separating the quadratic part $\hat{H}_{\rm G}$
is the essence of the Hartree approximation \cite{boyanovsky} or
the variational Gaussian approximation \cite{chang}. In fact, the
perturbation $\lambda \hat{H}_{\rm P}$, 
though $\hat{H}_{\rm G}$ contains a term of the same order as
the perturbation itself. Then the truncated quantum LvN equation
\begin{eqnarray}
i  \frac{\partial \hat{a}}{\partial t} + [ \hat{a},
\hat{H}_{\rm G}] = 0, \quad i \frac{\partial
\hat{a}^{\dagger}}{\partial t} + [ \hat{a}^{\dagger}, \hat{H}_{\rm
G}] = 0, \label{ln eq}
\end{eqnarray}
leads to the mean field equation
\begin{equation}
\ddot{u} (t) + \Bigl[ \pm \omega^2 + \frac{\lambda}{2} u^* u
\Bigr] u (t) = 0. \label{mean eq}
\end{equation}
The mean field equation above can also be obtained by minimizing
$E_{\rm G}$ in Eq. (\ref{vac en}), which proceeds by varying with
respect to $u^*$, using $\delta \dot{u}^* / \delta u^* = \partial/
\partial t$, and treating $u^*$ and $u$ independently. The equal-time
commutation relation now is guaranteed by the Wronskian condition
of Eq. (\ref{wron con}). The Gaussian vacuum state is annihilated
by $\hat{a}$
\begin{equation}
\hat{a} (t) \vert 0, t \rangle_{\rm G} = 0,
\end{equation}
and the excited number states are obtained by applying
$\hat{a}^{\dagger}$
\begin{equation}
\vert n, t \rangle_{\rm G} = \frac{\hat{a}^{\dagger
n}(t)}{\sqrt{n!}}  \vert 0, t \rangle_{\rm G}.
\end{equation}
These are the exact quantum states of the time-dependent
Schr\"{o}dinger equation only for $\hat{H}_{\rm G}$
\begin{equation}
i \frac{\partial}{\partial t} \vert n, t \rangle_{\rm G} =
\hat{H}_{\rm G} (t) \vert n, t \rangle_{\rm G}. \label{gaus eq}
\end{equation}
By fixing the time-dependent phase factor \cite{kim-kim} and
including the factor from $c$-number term in Eq. (\ref{2-p}), the
harmonic wave functions are given by
\begin{equation}
\Psi_{\rm G, n} (x) = \frac{1}{\sqrt{2^2 n!}} \Biggl(
\frac{1}{2 \pi u^* u } \Biggr)^{1/4} \Biggl(\frac{u}{\sqrt{u^* u}}
\Biggr)^{(2n+1)/2} H_n \Biggl(\frac{q}{\sqrt{2 u^* u}}
\Biggr)\exp \Biggl[\frac{i}{2} \frac{\dot{u}^*}{u^*}q^2 + i \frac{\lambda}{8}
\int^t (u^* u)^2 \Biggr]. \label{wav fn}
\end{equation}
where $H_n(x)$ are the Hermite polynomials and the last $c$-number term in 
the exponent comes from the corresponding $c$-number term in Eq. (\ref{2-p}). 
From now on we denote the wave functions (\ref{wav fn}) by $\vert n, t
\rangle_{\rm G}$ without any loss of generality. These states form
an orthonormal basis of the Fock space:
\begin{equation}
{}_{\rm G}\langle n, t \vert m, t \rangle_{\rm G} = \delta_{nm}.
\end{equation}

\subsection{Beyond the Hartree Approximation}

To go beyond the Hartree approximation, we need to include the
perturbation, which is now given by
\begin{equation}
\hat{H}_{\rm P} = \frac{1}{4!} \Bigl(u^{*4} \hat{a}^{\dagger 4} +
4 u^{*3} u \hat{a}^{\dagger 3} \hat{a} + 6 u^{*2} u^2
\hat{a}^{\dagger 2} \hat{a}^2 + 4 u^* u^3 \hat{a}^{\dagger}
\hat{a}^3 + u^4 \hat{a}^4 \Bigr). \label{pert}
\end{equation}
There was an attempt in Ref. \cite{bak} to solve Eq. (\ref{eq
aad}) for the full Hamiltonian including the perturbation
(\ref{pert}) by improving $\hat{a}$ and $\hat{a}^{\dagger}$. Here
we find the improved quantum states by directly solving the
Schr\"{o}dinger equation for the full Hamiltonian. The
perturbation excites and de-excites any Gaussian number state
$\vert n, t \rangle_{\rm G}$. As $\{ \vert n, t \rangle_{\rm G}
\}$ constitutes a Fock basis, we therefore expand the exact
quantum states as \cite{kim-khanna2}
\begin{equation}
\vert n, t \rangle = \sum_{l = 0}^{\infty} \sum_{m = 0}^{\infty}
\lambda^l C_{n; m}^{(l)} (t) \vert m, t \rangle_{\rm G},
\label{ser ex}
\end{equation}
where the lowest order coefficient is
\begin{equation}
C_{n; m}^{(0)} = \delta_{n,m}.
\end{equation}
Using the fact that any state, $\vert m, t \rangle_{\rm G}$
individually satisfies the Schr\"{o}dinger equation (\ref{gaus
eq}) for $\hat{H}_{\rm G}$, the Schr\"{o}dinger equation (\ref{sch
eq}) for the full Hamiltonian (\ref{anh osc}) leads to the set of
equations
\begin{equation}
\sum_{l = 0}^{\infty} \sum_{m = 0}^{\infty} i \lambda^l 
\dot{C}_{n; m}^{(l)} (t) \vert m, t \rangle_{\rm G} =
\sum_{l = 0}^{\infty} \sum_{m = 0}^{\infty} \lambda^l
C_{n; m}^{(l)} (t) \lambda  \hat{H}_{\rm P} \vert m, t
\rangle_{\rm G}.
\end{equation}
Comparing the powers of $\lambda$, we finally obtain a
hierarchy of dynamical equations for the coefficients
\begin{equation}
 \dot{C}_{n; m}^{(l)} (t) = - i  \sum_{j = 0}^{\infty}
 C_{n; j}^{(l-1)} (t) {}_G\langle m, t \vert
  \hat{H}_{\rm P} (t) \vert j, t \rangle_{\rm G}. \label{hier}
\end{equation}

Another expression for Eq. (\ref{ser ex}) may be obtained in a
compact form using operators $\hat{a}^{\dagger}$ and $\hat{a}$.
For each fixed $m$, we first sum over $l$
\begin{equation}
C_{n;m} (t) = \sum_{l = 0}^{\infty} \lambda^l
C_{n;m}^{(l)} (t),
\end{equation}
and then write any state, $\vert m, t \rangle_{\rm G}$ as either
an excited or a de-excited state of the given lowest order state
$\vert n, t \rangle_{\rm G}$, which is realized by applying the
creation or annihilation operators a certain number of times.
Hence, Eq. (\ref{ser ex}) can be written by introducing an
operator, $\hat{U}_1$ as
\begin{equation}
\vert n, t \rangle = \sum_{m = 0}^{\infty} C_{n;m} (t) \vert m, t
\rangle_{\rm G} \equiv \hat{U}_{\rm I} [\hat{a}^{\dagger} (t),
\hat{a} (t); t, \lambda] \vert n, t \rangle_{\rm G},
\end{equation}
and the Schr\"{o}dinger equation leads to
\begin{eqnarray}
\Biggl[i \frac{\partial}{\partial t} \hat{U}_{\rm I} (t,
\lambda) + [\hat{U}_{\rm I} (t, \lambda), \hat{H}_{\rm G}] -
\lambda \hat{H}_{\rm P} \Biggr] \vert n, t \rangle_{\rm G}
= 0. \nonumber
\end{eqnarray}
Using Eq. (\ref{ln eq}) and technically assuming that all time
derivatives act only on $c$-numbers but not on the operators
$\hat{a}^{\dagger}$ and $\hat{a}$, we obtain an interaction
picture-like equation for the operator $\hat{U}$:
\begin{equation}
i \frac{\partial}{\partial t} \hat{U}_{\rm I} (t, \lambda) =
\lambda \hat{H}_{\rm P} \hat{U}_{\rm I} (t, \lambda).
\label{int op}
\end{equation}
We then obtain the formal solution
\begin{eqnarray}
\vert n, t \rangle = \hat{U}_{\rm I} (t, \lambda) \vert n, t
\rangle_{\rm G}, ~~ (n = 0, 1, 2, \cdots),  \label{op ex}
\end{eqnarray}
where
\begin{equation}
\hat{U}_{\rm I} (t, \lambda) = {\rm T} \exp\Biggl[- i \lambda
\int \hat{H}_{\rm P} dt \Biggr]. \label{form sol}
\end{equation}
Here T denotes a time-ordering for the integral and
$\hat{a}^{\dagger}(t)$ and $\hat{a}(t)$ are treated as if they are
constant operators.

We now find the improved vacuum state up to any order either by
solving Eq. (\ref{hier}) or by acting with the operator in Eq.
(\ref{form sol}) on the Gaussian vacuum state. For instance, the
improved vacuum state to order $\lambda^2$ is given by
\begin{equation}
\vert 0, t \rangle_{[2]} = \vert 0, t \rangle_{\rm G} + \lambda
 \sum_{m = 0} C_{0;m}^{(1)} (t) \vert m, t \rangle_{\rm G} +
\lambda^2 \sum_{m = 0} C_{0;m}^{(2)} (t) \vert m, t
\rangle_{\rm G}, \label{ng vac}
\end{equation}
where the only nonvanishing coefficients are
\begin{equation}
C_{0;4}^{(1)} (t) = - i \frac{1}{\sqrt{4!}}\int^t u^{*4} (t'),
\label{fir co}
\end{equation}
and
\begin{eqnarray}
C_{0;8}^{(2)} (t) &=& (- i)^2 \frac{\sqrt{70}}{4!} \int^t u^{*4}
(t') \int^{t'} u^{*4} (t''), \nonumber\\ C_{0;6}^{(2)} (t) &=& (-
i)^2 \frac{\sqrt{5}}{3} \int^t u^{*3} (t') u(t') \int^{t'} u^{*4}
(t''), \nonumber\\ C_{0;4}^{(2)} (t) &=& (- i)^2
\frac{3}{\sqrt{4!}} \int^t u^{*2} (t') u^2 (t') \int^{t'} u^{*4}
(t''), \nonumber\\ C_{0;2}^{(2)} (t) &=& (- i)^2 \frac{1}{3
\sqrt{2}} \int^t u^{*} (t') u^3 (t')  \int^{t'} u^{*4} (t''),
\nonumber\\ C_{0;0}^{(2)} (t) &=& (- i)^2 \frac{1}{4!} \int^t
u^{4} (t') \int^{t'} u^{*4} (t''). \label{sec co}
\end{eqnarray}
The non-Gaussian nature of the vacuum state (\ref{ng vac}) can be
exploited by calculating the kurtosis (higher moments). The
two-point and four-point correlators with respect to the Gaussian
vacuum state (\ref{wav fn}) are
\begin{eqnarray}
{}_{\rm G}\langle 0, t \vert \hat{x}^2 \vert n, t \rangle_{\rm G}
&=& u^* u, \nonumber\\ {}_{\rm G}\langle 0, t \vert
\hat{x}^4 \vert 0, t \rangle_{\rm G} &=& 3 (u^*u)^2,
\end{eqnarray}
whereas those with respect to the improved vacuum state (\ref{ng
vac}) are given by
\begin{eqnarray}
{}_{[2]}\langle 0, t \vert \hat{x}^2 \vert 0, t \rangle_{[2]} &=&
 u^*u + \lambda^2 [\sqrt{2}(C^{(2)}_{0;2}u^2 +
C^{(2)*}_{0;2} u^{*2}) + (C^{(2)*}_{0;0} + C^{(2)}_{0;0} + 9
C^{(1)*}_{0;4} C^{(1)}_{0;4}) u^* u ] + {\cal O} (\lambda^3), 
\label{2 fn}\\ {}_{[2]}\langle 0, t \vert \hat{x}^4
\vert 0, t \rangle_{[2]} &=&  3(u^*u)^2 + \sqrt{4!}
\lambda (C_{0;4}^{(1)} u^4 + C_{0;4}^{(1)*}u^{*4} ) +
\lambda^2 [\sqrt{4!}(C^{(2)}_{0;4}u^4 + C^{(2)*}_{0;4}u^{*4}) 
\nonumber\\ &+& 
6\sqrt{2}(C^{(2)}_{0;2} (u^* u) u^2 + C^{(2)*}_{0;2} (u^* u)
u^{*2}) + (123 C^{(1)*}_{0;4} C^{(1)}_{0;4} + 3 C^{(2)*}_{0;0} +
3C^{(2)}_{0;0}) (u^* u)^2] + {\cal O} (\lambda^3).
\label{4 fn}
\end{eqnarray}

\subsection{Stability of the LvN Method}

The stability of the LvN method should be checked since it is a
time-dependent perturbation theory. That is, any secular
coefficient in Eq. (\ref{ser ex}) should be removed systematically
to insure the physically meaningful solution. For that purpose, we
compare the LvN method with the standard perturbation theory for
the well-known anharmonic oscillator with unbroken symmetry
(positive sign in Eq. (\ref{anh osc})). In that case we find the
solution to the auxiliary mean-field equation (\ref{mean eq})
\begin{equation}
u (t) = \frac{1}{\sqrt{2 \Omega}} e^{- i \Omega t}, \label{sol}
\end{equation}
where $\Omega$ is given by the gap equation
\begin{equation}
\Omega^2 = \omega^2 + \frac{\lambda}{4 \Omega}.
\end{equation}
Then the time-dependent wave function (\ref{wav fn}) is given by
\begin{eqnarray}
\Psi_{\rm G, n} (x, t) &=& \exp \Biggl[- i \Biggl( \Omega ( n +
1/2 )- \frac{\lambda}{32 \Omega^2} \Biggr) t \Biggr] \times
\frac{1}{\sqrt{2^2 n!}} \Biggl(\frac{\Omega}{\pi}
\Biggr)^{1/4} H_n \Biggl(\sqrt{\Omega} q \Biggr)
\exp \Biggl[- \frac{\Omega}{2} q^2 \Biggr] \nonumber\\ &=&
\exp \Biggl[- i \Biggl( \Omega (n + 1/2)- \frac{\lambda}{32
\Omega^2} \Biggr) t \Biggr] \times \Psi_{\rm G, n} (x),
\end{eqnarray}
where $\Psi_{\rm G,n} (x)$ denotes the harmonic oscillator wave
function.

Substituting the solution (\ref{sol}) into Eqs. (\ref{fir co}) and
(\ref{sec co}), we find the improved vacuum state corrected to
${\cal O}(\lambda^2)$:
\begin{eqnarray}
\vert 0, t \rangle_{[2]} &=& \exp \Biggl[- i \Biggl(
\frac{\Omega}{2} - \frac{\lambda}{32 \Omega^2} \Biggr) t
\Biggr] \Biggl[ \Biggl(1 + i \frac{\lambda^2}{2^9 \cdot 3
\Omega^5} t \Biggr) \vert 0 \rangle_{\rm G} + 
\frac{\lambda^2 }{2^{7} \cdot 3 \sqrt{2} \Omega^6} \vert 2 \rangle_{\rm
G} \nonumber\\ &-& \Biggl(  \frac{\lambda}{2^5\sqrt{6}
\Omega^3} - \frac{\sqrt{3} \lambda^2}{2^{9} \sqrt{2}
\Omega^6} \Biggr) \vert 4 \rangle_{\rm G} 
+ \frac{\lambda^2}{2^{7} \cdot 3^2 \Omega^6} \vert 6 \rangle_{\rm G} +
\frac{\sqrt{70}\lambda^2 }{2^{12} \cdot 3 \Omega^6} \vert
8 \rangle_{\rm G} \Biggr] + {\cal O} (\lambda^3).
\end{eqnarray}
Note that $C^{(2)}_{0;0}$ originating from the four quanta
creation and the subsequent annihilation leads to a secular term
increasing as $t$. This is not a drawback of the LvN method but
just a consequence of the time-dependent perturbation theory
searching time-dependent states. As the coefficient of $\vert 0, t
\rangle_{\rm G}$ is the first two terms of 
$\exp [i \lambda^2/(2^9 \cdot 3 \Omega^5) t ]$, it can be approximately
absorbed into the overall time-dependent factor to ${\cal O}(\lambda^2)$
\begin{equation}
\exp \Biggl[- i \Biggl( \frac{\Omega}{2} - \frac{\lambda}{32
\Omega^2}  - \frac{\lambda^2}{2^9 \cdot 3 \Omega^5}
\Biggr) t \Biggr]. \label{phase}
\end{equation}
This factor coincides with the time-dependent solution to the full
Schr\"{o}dinger equation obtained from time-independent
perturbation method \cite{kleinert}, where the corrected energy at
${\cal O}(\lambda^2)$ is
\begin{equation}
E_{[2]} = \frac{\Omega}{2} - \frac{\lambda}{32
\Omega^2} - \frac{\lambda^2}{2^9 \cdot 3 \Omega^5}
+ {\cal O} (\lambda^3).
\end{equation}
The higher order terms that come from the creation of even number
of quanta and its subsequent annihilation of equal quanta or vise
versa also contain secular terms proportional to powers of $t$
depending on the number of such processes. All these terms will
provide the correct energy to the Schr\"{o}dinger equation.

Another way to understand this phase factor and thereby secular
terms is to use the formal solution (\ref{form sol}). The operator
has an exponential form \cite{kim5}
\begin{equation}
\hat{U}_{\rm I} (t, \lambda) = \exp \Biggl[ - i \lambda
\int^t \hat{H}_{\rm P} (t') + (-i \lambda)^2 [\int^t dt'
\hat{H}(t'), \int^{t'} dt'' \hat{H}_{\rm P} (t'')] + {\cal O}
(\lambda^3) \Biggr].
\end{equation}
A $c$-number term from the commutator
\begin{equation}
\exp\Biggl[- \frac{\lambda^2}{(4!)^2} \int^t dt' u^4 (t')
\int^{t'} dt'' u^{*4} (t'') [ \hat{a}^4, \hat{a}^{\dagger 4}]
\Biggr] \rightarrow \exp\Biggl[ i \frac{\lambda^2}{2^9 \cdot 3 \Omega^5}
t \Biggr]
\end{equation}
is nothing but the phase factor (\ref{phase}). Now the
time-dependent vacuum state to ${\cal O}(\lambda^2)$ {\it does
not} involve any secular term as shown
\begin{eqnarray}
\vert 0, t \rangle_{[2]} &=& e^{i E_{[2]} t}
\Biggl\{ \vert 0 \rangle_{\rm G} + e^{- i 
\frac{\lambda^2}{2^9 \cdot 3 \Omega^5} t} \Biggl[ \frac{\lambda^2
}{2^{7} \cdot 3 \sqrt{2} \Omega^6} \vert 2 \rangle_{\rm G} -
\Biggl( \frac{\lambda}{2^5\sqrt{6} \Omega^3} -
\frac{\sqrt{3} \lambda^2}{2^{9} \sqrt{2} \Omega^6} \Biggr)
\vert 4 \rangle_{\rm G} \nonumber\\ ~~~~&+& 
\frac{\lambda^2}{2^{7} \cdot 3^2 \Omega^6} \vert 6 \rangle_{\rm G} +
\frac{\sqrt{70}\lambda^2 }{2^{12} \cdot 3 \Omega^6} \vert
8 \rangle_{\rm G} \Biggr] \Biggr\} + {\cal O} (\lambda^3).
\end{eqnarray}

We thus have shown that the seemingly secular behavior can be
removed systematically by taking the proper time-dependent phase
factor for the wave function. This phase factor yields the correct
energy for the anharmonic oscillator (\ref{anh osc}) to any
desired order. The idea of removing the secular terms of higher
order corrections by absorbing them into the corrected energy is
equivalent to removing the secular terms by renormalizing
frequency in the multiple-scale perturbation theory \cite{bender}.
We note that the LvN method proves very accurate because the
lowest order vacuum state is a Gaussian state that extremizes the
Hamiltonian and the corrected vacuum state is expanded in the Fock
basis \cite{kleinert}. Further, the LvN method is a powerful tool to
finding the quantum states for nonequilibrium systems that are
undergoing phase transitions \cite{kim-lee2}. For the field theory 
case, the elimination of secular terms can be achieved using 
multiple-scale perturbation theory and this aspect has been discussed 
in detail in \cite{ksk}.

There is another kind of instability originating from the dynamics
of the system itself. Phase transitions from an explicit symmetry
breaking in time provide such a dynamical instability. The
anharmonic oscillator (\ref{anh osc}) whose quadratic potential
changes signs from positive to negative can be a quantum
mechanical analog for phase transition. After the symmetry is
broken, the mean field equation
\begin{equation}
\ddot{u} + \Bigl[- \omega^2 + \frac{\lambda}{2} u^* u \Bigr]
u = 0
\end{equation}
may have a period when the quadratic term $(- \omega^2)$ dominates
over the quartic one $(\lambda u^* u/2)$. Then the $u$ grows
exponentially as $u \approx e^{\omega t}/\sqrt{2 \omega}$ until
the quartic term grows and becomes comparable to the quadratic
one. During this period all higher order corrections $C^{(l)}$
beyond the Hartree approximation grow exponentially as powers of
$u$ and $u^*$. This dynamical instability ceases when the state
reaches the true vacuum state and oscillates over it. Thus the
dynamical instability for a limited period does not cause any
serious secular behavior as for the static system.

\section{$\phi^4$-Field Theory in the Hartree Approximation}

Now we will apply the LvN method to the $\phi^4$-field theory, but
first work out correlation functions within the Hartree
approximation in this section. The $\phi^4$-field theory to be
considered in this paper has the Hamiltonian in D-space dimensions
\begin{equation}
\hat{H} (t) = \int d^D x \Biggl[\frac{1}{2}\hat{\pi}^2 +
\frac{1}{2} (\nabla \hat{\phi})^2 + \frac{m^2}{2} \hat{\phi}^2 +
\frac{\lambda}{4!} \hat{\phi}^4 \Biggr], \label{eq free}
\end{equation}
where $\hat{\pi}(x) = \hat{\dot{\phi}}(x)$ is the conjugate
momentum operator. We divide the quantum field $\hat{\phi}$ into a
classical background and quantum fluctuations over this
background, $\hat{\phi}(x) = \phi_{c}(x) + \hat{\phi}_{f}(x)$,
where the classical  background (mean) field $\phi_{c}(x)$ is in
general considered to be spatially homogeneous. Then the
Hamiltonian can be decomposed as
\begin{equation}
\hat{H}(t) = H_{\rm c}(t) + \hat{H}_{\rm f}(t) + \hat{H}_{\rm
int}(t) + \delta\hat{H}_{\rm int}(t) \label{ham sum}
\end{equation}
where
\begin{eqnarray}
H_{\rm c}(t) &=& \int d^D x \Biggl[\frac{1}{2}\pi_{c}^2 +
\frac{1}{2} (\nabla \phi_{c})^2 + \frac{m^2}{2} \phi_{c}^2 +
\frac{\lambda}{4!}\phi_{c}^4 \Biggr], \nonumber\\ \hat{H}_{\rm
f}(t) &=& \int d^D x \Biggl[\frac{1}{2} \hat{\pi_{f}}^2+
\frac{1}{2}(\nabla\hat{\phi}_{f})^2 + \frac{m^2}{2}
\hat{\phi}_{f}^2 + \frac{\lambda}{4!}\hat{\phi}_{f}^4 \Biggr],
\nonumber \\ \hat{H}_{\rm int}(t) &=& \int d^D x \Biggl[
\frac{\lambda}{4} \phi_{c}^2 \hat{\phi}_{f}^2 \Biggr],
 \nonumber \\
\delta\hat{H}_{\rm int}(t) &=& \int d^D x \Biggl[ \pi_{c}
\hat{\pi}_{f} + m^2\phi_{c} \hat{\phi}_{f} +
\frac{\lambda}{3!}\phi_{c}(\phi_{c}^2 +
\hat{\phi}_{f}^2)\hat{\phi}_{f} \Biggr].
\end{eqnarray}
Here $H_{\rm c}$ and $\hat{H}_{\rm f}$ are purely classical and
quantum parts. We have divided the interaction into $\hat{H}_{\rm
int}$ with even powers of $\hat{\phi}_{f}$ and $\delta
\hat{H}_{\rm int}$ with odd powers of $\hat{\pi}_f$ and
$\hat{\phi}_f$. The $\phi_{c}(x)$ is nonvanishing only for the
case when symmetry is spontaneously broken. We treat individually
and collectively the  modes in the momentum space of quantum
fluctuations and the inhomogeneous background field. The momentum
modes are given by the Fourier transform of a field (operator) and
its inverse transform
\begin{eqnarray}
F (x) &=& \int [d k] F (k)e^{i k \cdot x}, \nonumber
\\ F (k) &=& \int d^D x F ( x) e^{- i
k \cdot x},
\end{eqnarray}
where $F$ denotes either $\hat{\phi}_{f} (x)$ and $\hat{\pi}_{f}
(x)$ or $\phi_{c} (x)$ and $\pi_{c}(x)$, and
\begin{equation}
[d k] = \frac{d^D k}{(2 \pi)^D}.
\end{equation}
Then the quadratic integral takes the form
\begin{equation}
\int d^D x \hat{F}^2 (x) = \int [d k] \hat{F}(k) \hat{F} (- k).
\end{equation}
The Fourier modes of $\hat{\phi}_f$ and $\hat{\pi}_f$ will be
denoted by
\begin{eqnarray}
\hat{\Phi}_{k} &=& \int d^D x \hat{\phi}_f (x) e^{- i k \cdot x },
\nonumber \\ \hat{\Pi}_{k} &=& \int d^D x \hat{\pi}_f (x) e^{ - i
 k \cdot x}.
\end{eqnarray}
The hermiticity of $\hat{\phi}_f$ and $\hat{\pi}_f$ implies that
$\hat{\Phi}_{k}^{\dagger} = \hat{\Phi}_{- k}$ and $\hat{\Pi}_{
k}^{\dagger} = \hat{\Pi}_{- k}$.

The commutation relation of the fields
\begin{equation}
[\hat{\phi}_f (x), \hat{\pi}_{f} (y) ] = i \delta ({\bf x}-
{\bf y}),
\end{equation}
leads to those of modes in the momentum space
\begin{equation}
[\hat{\Phi}_{{k}'}, \hat{\Pi}_{ k} ] = i (2 \pi)^D \delta
(k' + k).
\end{equation}
In terms of the annihilation and creation operators satisfying the
equal time commutation relation
\begin{equation}
[\hat{a}_{k'}(t), \hat{a}^{\dagger}_{k} (t) ] = (2 \pi)^D \delta
(k'- k), \label{mode com}
\end{equation}
the momentum space operators $\hat{\Phi}_{k}$ and $\hat{\Pi}_{k}$
may be expressed as
\begin{eqnarray}
\hat{\Phi}_{k} = \varphi_{k}(t) \hat{a}_{k}(t) +
\varphi_{-k}^{*}(t) \hat{a}_{-k}^{\dagger}(t), \nonumber
\\ \hat{\Pi}_{k} = \dot{\varphi}_{k}(t) \hat{a}_{k} (t)
+ \dot{\varphi}_{-k}^{*}(t) \hat{a}_{-k}^{\dagger} (t).
\label{osc rep}
\end{eqnarray}
Here it is assumed that $\varphi_{-k}(t) = \varphi_{k}(t)$ and
\begin{equation}
\dot{\varphi}_{k}^* (t) \varphi_{k}(t)  - \varphi_{k}^*(t)
\dot{\varphi}_{k}(t) = i.
\end{equation}
Then the annihilation and creation operators are also expressed as
\begin{eqnarray}
\hat{a}_{k} (t) &=& i [\varphi_{k}^*(t)
\hat{\Pi}_{k} - \dot{\varphi}_{k}^*(t) \hat{\Phi}_{k}],
\nonumber\\ \hat{a}_{k}^{\dagger} (t) &=& - i
[\varphi_{-k}(t) \hat{\Pi}_{-k} - \dot{\varphi}_{-k}(t)
\hat{\Phi}_{-k}].
\end{eqnarray}
Note that the momentum space operators $\hat{\Phi}_{k}$ and
$\hat{\Pi}_{k}$ are regarded as time-independent ones whereas
$\hat{a}_{k} (t)$ and $\hat{a}_{k}^{\dagger} (t)$ as
time-dependent Liouville ones in the LvN picture. The Gaussian
vacuum is the state annihilated by all $\hat{a}_{k} (t)$:
\begin{equation}
\hat{a}_{k} (t) \vert 0, t \rangle_{\rm G} = 0,
\end{equation}
or the product of the Gaussian vacuum state for each $\hat{a}_{k}
(t)$
\begin{equation}
\vert 0, t \rangle_{\rm G} = \prod_{k} \vert 0_{k}, t \rangle_{\rm
G}.
\end{equation}

In the Hartree approximation we consider only those quadratic
terms from $\hat{H}_{\rm f}$ and $\hat{H}_{\rm int}$ whose
Gaussian vacuum expectation values do not vanish. Then the
interaction term has the Fourier modes
\begin{eqnarray}
\hat{H}_{\rm int} &=& \frac{\lambda}{4} \int [d k_1] \int [d k_2]
\int [d k_3] \int [d k_4] (2 \pi)^D \delta (k_1 + k_2 + k_3 + k_4)
\phi_c (k_1) \phi_c (k_2) \hat{\Phi}_{k_3} \hat{\Phi}_{k_4}
\nonumber\\ &\rightarrow& \frac{\lambda}{4} \int [d k_1] \int [d
k_2] (2 \pi)^D \delta (k_1 + k_2) \phi_c (k_1) \phi_c (k_2) \int
[d k_3] \hat{\Phi}_{k_3} \hat{\Phi}_{- k_3} \nonumber\\ &=&
\frac{\lambda}{4} \phi_c^2 (x) \int [d k] \hat{\Phi}_{k}
\hat{\Phi}_{- k}.
\end{eqnarray}
The quartic term in the Hamiltonian for the fluctuation field
$\phi_f$ can be approximated as
\begin{eqnarray}
\hat{\phi}_{f}^4 \rightarrow 6 \langle \hat{\phi}_{f}^2
\rangle_{\rm G} \hat{\phi}_{f}^2, \label{eq ha}
\end{eqnarray}
Then the resulting quadratic part takes the form
\begin{equation}
\hat{H}_{\rm G} = \int [d k] \Biggl[ \frac{1}{2} \hat{\Pi}_{k}
\hat{\Pi}_{-k} + \frac{1}{2} \Biggl(\omega_{k}^2 +
\frac{\lambda}{2} \phi_c^2 + \frac{\lambda}{2} \int [d k'] \langle
\hat{\Phi}_{k'} \hat{\Phi}_{-k'} \rangle_{\rm G} \Biggr)
\hat{\Phi}_{k} \hat{\Phi}_{-k} \Biggr], \quad (\omega_{k}^2 = m^2
+ k^2).
\end{equation}
Under the field redefinition $\hat{\Pi}_{\pm k} = (\hat{\Pi}_{k}
\pm \hat{\Pi}_{-k})/2$ and $\hat{\Phi}_{\pm k} = (\hat{\Phi}_{ k}
\pm \hat{\Phi}_{- k})/2$, this Hamiltonian is equivalent to that
of harmonic oscillators
\begin{equation}
\hat{H}_{\rm G}(t) = \sum_{\kappa} \Biggl[  \frac{1}{2}
\hat{\Pi}_{\kappa}^2(t) + \frac{1}{2} \Omega_{\kappa}^2 (t)
\hat{\Phi}_{\kappa}^2 (t) \Biggr], \label{eq ham}
\end{equation}
where
\begin{equation}
\Omega_{\kappa}^2 (t) =  \omega_{k}^2 + \frac{\lambda}{2} \phi_c^2
(x) + \frac{\lambda}{2} \int [d k'] \langle \hat{\Phi}_{k'}
\hat{\Phi}_{-k'} \rangle_{\rm G}. \label{omega}
\end{equation}
We may identify the classical background $\phi_c$ and $\pi_c$
either with the vacuum expectation values of $\hat{\phi}$ and
$\hat{\pi}$ or with the coherent state expectation values of
$\hat{\phi}_f$ and $\hat{\pi}_f$, respectively:
\begin{eqnarray}
\phi_{c}(x)  &=& \langle \hat{\phi} (x) \rangle_{\rm vac} =
\langle \hat{\phi}_f (x) \rangle_{\rm cs} \equiv \int [d k]
\langle \hat{\Phi}_{k} \rangle_{\rm cs} e^{i k \cdot x}, \nonumber
\\ \pi_{c}(x) &=& \langle \hat{\pi} (x)
\rangle_{\rm vac} =  \langle \hat{\pi}_f (x) \rangle_{\rm cs}
\equiv \int [d k] \langle \hat{\Pi}_{k} (t) \rangle_{\rm cs} e^{i
 k \cdot x}, \label{eq fcs}
\end{eqnarray}
where $\langle \cdots \rangle_{\rm cs}$ denotes the expectation
value taken with respect to the coherent state:
\begin{eqnarray}
\langle \hat{\Phi}_{k} \rangle_{\rm cs} &=& 
\alpha_{k} \varphi_{k}(t) + \alpha_{- k}^{*} \varphi_{-
k}^{*}(t), \nonumber \\ \langle \hat{\Pi}_{k} \rangle_{\rm cs}
&=& \alpha_{k}\dot{\varphi}_{k}(t) + \alpha_{-
k}^{*}\dot{\varphi}_{- k}^{*}(t).
\end{eqnarray}

Now the equations for the auxiliary field variables $\varphi_{k}$
and $\varphi^{*}_{k}$ can be easily obtained in the Hartree
approximation by making use of the LvN equations for $\hat{a}_{k}
(t)$ and $\hat{a}_{k}^{\dagger} (t)$ for the Hamiltonian (\ref{eq
ham}). Then the LvN equations
\begin{eqnarray}
i \frac{\partial \hat{a}^{\dagger}_{k} (t)}{\partial t} +
[\hat{a}^{\dagger}_{k}(t) , \hat{H}_{\rm G} (t)] &=& 0, \nonumber
\\ i  \frac{\partial \hat{a}_{k}(t)}{\partial t} +
[\hat{a}_{k} (t) , \hat{H}_{\rm G} (t)] &=& 0, \label{eq lvn}
\end{eqnarray}
lead to the equations
\begin{eqnarray}
\ddot{\varphi}_{k}(t) + \Biggl[\omega_{k}^2 +
\frac{\lambda}{2}\phi_{c}^2 + \frac{\lambda}{2} \Biggl(\int
[d k'] \varphi_{k'}^{*}\varphi_{k'} \Biggr) \Biggr] \varphi_{
k}(t) &=& 0, \nonumber
\\ \ddot{\varphi}_{k}^{*}(t) + \Biggl[\omega_{k}^2 +
\frac{\lambda}{2}\phi_{c}^2 + \frac{\lambda}{2} \Biggl(\int
[d k'] \varphi_{k'}^{*}\varphi_{k'} \Biggr) \Biggr] \varphi_{
k}^{*}(t) &=& 0. \label{eq ph}
\end{eqnarray}
The equation for the classical background $\phi_{c}(t)$ is
obtained from the effective classical Hamiltonian, $H_{\rm c} (t)
+ \langle \hat{H}_{\rm int} (t) \rangle_{\rm G}$, as
\begin{equation}
\ddot{\phi}_{c} ({\bf x}, t) - \nabla^2 \phi_c ({\bf x}, t) +
\Biggl[ m^2 + \frac{\lambda}{3!}\phi_{c}^2 ({\bf x}, t) +
\frac{\lambda}{2} \Biggl(\int [d^{D}k']\varphi_{k^{\prime}}^{*}
\varphi_{k'} \Biggr)\Biggr]\phi_{c}({\bf x},t) = 0.
\end{equation}

It is more advantageous to work with the evolution equations for
correlation functions, rather than the field equations. Firstly,
the 2-point correlation functions in thermal equilibrium are
related to the Bose-Einstein distribution function (for a quantum
theory) and the temperature of the system (for a classical theory)
in a fairly simple way; thereby allowing us to use them as
benchmarks to track the evolution of the system towards thermal
equilibrium. Secondly, it is more convenient to  make systematic
improvements to the mean field description, by working with
equations for the correlation functions. As we shall see in
Section VI, the mean field equations for the 2-point correlation
functions are part of an infinite hierarchy of evolution equations
for the connected, equal-time, n-point correlators and are
obtained by truncating this hierarchy at the level of the 2-point
functions.

We define the 2-point subtracted correlation functions for the
fields as
\begin{eqnarray}
g_{11}(x, x';t) &\equiv& \langle [\hat{\phi}_f (x) -
\phi_{c}(x)][\hat{\phi}_f (x') - \phi_{c} (x')] \rangle_{{\rm cs}}
\equiv \langle\hat{\phi}_f(x)\hat{\phi}_f (x')\rangle_{{\rm v}}
\nonumber
\\ g_{22}(x, x';t) &\equiv& \langle
[\hat{\pi}_f (x) - \pi_{c}(x)][\hat{\pi}_f (x') - \pi_{c} (x')]
\rangle_{{\rm cs}} \equiv \langle\hat{\pi}_f(x)\hat{\pi}_f(
x')\rangle_{{\rm v}} \nonumber\\
 g_{12}(x, x';t) &\equiv& \langle [\hat{\phi}_f
(x) - \phi_{c}(x)][\hat{\pi}_f (x') - \pi_{c} (x')] \rangle_{{\rm
cs}} \equiv \langle\hat{\phi}_f(x)\hat{\pi}_f (x)\rangle_{{\rm v}}
\nonumber\\ g_{21}(x, x';t) &\equiv& \langle [\hat{\pi}_f (x) -
\pi_{c}(x)][\hat{\phi}_f (x') - \phi_{c} (x')] \rangle_{{\rm cs}}
\equiv \langle\hat{\pi}_f(x)\hat{\phi}_f (x')\rangle_{{\rm v}}
\end{eqnarray}
The subscript `v' implies expectation value with respect to the vacuum state.
Using the above expressions we get
\begin{eqnarray}
g_{11}(x, x';t) &=& \int [d k] \varphi^{*}_{k} (t)
\varphi_{k}(t) e^{i k \cdot( x - x')}, \nonumber
\\ g_{22}(x,x';t) &=& \int [d k]
 \dot{\varphi}^{*}_{k}(t) \dot{\varphi}_{k} (t) e^{i k
\cdot(x - x')}, \nonumber \\ g_{12}(x, x';t) &=& \int [d k]
\dot{\varphi}^{*}_{k} (t) \varphi_{k}(t) e^{i k \cdot(x - x')},
\nonumber
\\ g_{21}(x, x';t) &=& \int [d k]
 \varphi^{*}_{k} (t) \dot{\varphi}_{k} (t) e^{i k \cdot (x -
x')}. \label{eq cfl}
\end{eqnarray}
From Eq. (\ref{eq cfl}), the equal-time 2-point correlation
functions $G_{ij}(k,t)$ in momentum space can be defined through
the Fourier transform
\begin{equation}
g_{ij}(x, x';t) = \int [d k] G_{ij}(k,t)e^{i k \cdot(x - x')},
\label{eq GK}
\end{equation}
where $i,j=1,2$. Taking the time-derivative of the 2-point
correlation functions yields the following evolution equations
\begin{eqnarray}
\dot{g}_{11}(x, x';t) &=& g_{12}(x, x';t) + g_{21}(x, x';t)
\nonumber \\ \dot{g}_{22}(x, x';t) &=& \int [d k]
[\ddot{\varphi}_{k}^{*} (t) \dot{\varphi}_{k} (t) +
\dot{\varphi}_{k}^{*}(t) \ddot{\varphi}_{k} (t)] e^{i k \cdot (x -
x')} \label{eq evl} \\ \dot{g}_{12}(x, x';t) &=& \int [d k]
[\ddot{\varphi}_{k}^{*} (t) \varphi_{k} (t) +
\dot{\varphi}_{k}^{*} (t) \dot{\varphi}_{k} (t)] e^{i k \cdot (x -
x')} \nonumber
\end{eqnarray}

Substituting the expressions for $\ddot{\varphi}_{k}$ and
$\ddot{\varphi}_{k}^{*}$ in Eq. (\ref{eq evl}) and making use of
Eq. (\ref{eq GK}) results in the following evolution equations for
the correlation functions in the momentum space in the Hartree
approximation up to ${\cal O}(\lambda^2)$:
\begin{eqnarray}
\dot{G}_{11}(k,t) &=& G_{12}(k,t) + G_{21}(k,t), \nonumber
\\ \dot{G}_{22}(k,t) &=& - \Bigl[\omega_{k}^2 +
\frac{\lambda}{2}\phi_{c}^2(t) + \frac{\lambda}{2}g_{11}(0 ,t)
\Bigr](G_{12}(k,t) + G_{21}(k,t)) + {\cal O}(\lambda^2),
\nonumber
\\ \dot{G}_{12}(k,t) &=& G_{22}(k,t) -
\Bigl[\omega_k^2 + \frac{\lambda}{2}\phi_{c}^2(t) +
\frac{\lambda}{2}g_{11}(0,t) \Bigr]G_{11}(k,t) + {\cal
O}(\lambda^2). \label {eq gcor}
\end{eqnarray}
As shown in \cite{wett,aarts}, one can define a quantity
$\gamma^{-2}(k)$ in terms of the momentum space 2-point
correlation functions, which is found to be conserved in the
Hartree approximation
\begin{eqnarray}
\gamma^{-2}(k) &=& G_{11}(k,t)G_{22}(k,t) -
\frac{1}{4}[G_{12}(k,t) + G_{21}(k,t)]^2 \nonumber\\ &=& -
\frac{1}{4} (\dot{\varphi}_{k}^* \varphi_{k} - \varphi_{
k}^* \dot{\varphi}_{k})^2 = \frac{1}{4}
\end{eqnarray}
as a direct consequence of the Wronskian condition, Eq. (\ref{wron
con}). Issues of renormalization of this model have been
investigated in detail in Ref. \cite{ssg2}.

It is also possible to define {\it thermal} 2-point functions by
taking the expectation value of the field operators with respect
to an initial Gaussian thermal  state
\begin{equation}
\langle \hat{\phi}(x,t)\hat{\phi}(x',t)\rangle _{\rm T} =
\frac{1}{{\cal Z}}{\rm Tr}[e^{-\beta_{0}\hat{H}_{\rm G}}
\hat{\phi}_f(x)\hat{\phi}_f(x')]
\end{equation}
where $\beta_{0}$ is the initial inverse temperature of the system
and
\begin{equation}
{\cal Z} \equiv {\rm Tr}[e^{-\beta_{0}\hat{H}_{\rm G}}] =
\sum_{n_{k}=0}^{\infty} \langle n_k,t \vert
e^{-\beta_{0} \Omega_{k}(\hat{a}_{k}^{\dagger} \hat{a}_{
k}+\frac{1}{2})} \vert n_k,t \rangle.
\end{equation}
The corresponding equations for the auxiliary field variables, are
again obtained from the LvN equations using the Hamiltonian
(\ref{eq ham}) but with the expectation values taken with respect
to the initial {\it thermal} Gaussian ensemble. This leads to the
set of equations
\begin{eqnarray}
\ddot{\varphi}_{k}(t) + \Biggl[\omega_{k}^2 +
\frac{\lambda}{2}\phi_{c}^2 + \frac{\lambda}{2} \Biggl(\int
[d k] (2n_{k'}+1) \varphi_{k'}^{*}\varphi_{k'}\Biggr)\Biggr]
\varphi_{k}(t) &=& 0, \nonumber \\ \ddot{\varphi}_{k}^{*}(t) +
\Biggl[\omega_{k}^2 + \frac{\lambda}{2}\phi_{c}^2 +
\frac{\lambda}{2} \Biggl(\int [d k](2n_{k'} + 1)
\varphi_{k'}^{*}\varphi_{k'} \Biggr)\Biggr] \varphi_{k}^{*}(t) &=&
0, \nonumber \\ \ddot{\phi}_{c} (x, t) - \nabla^2 \phi_c (x, t) +
\Biggl[m^2 + \frac{\lambda}{3!}\phi_{c}^2 (x, t) 
+ \frac{\lambda}{2} \Biggl(\int [d k](2n_{k'}+1) \varphi_{
k'}^{*}\varphi_{k'} \Biggr) \Biggr]\phi_{c}(x, t) &=& 0, \label{eq
tfh}
\end{eqnarray}
where $n_{k} (T)$ is the Bose-Einstein distribution for a field
theory system at a temperature $T= 1/\beta_0$:
\begin{equation}
n_{k}(T) = \frac{1}{e^{\beta_0 \Omega_{k}}-1}.
\end{equation}
Note that at $T=0$, the above set of equations reduces to Eqs.
(\ref{eq ph}) and (102). The expressions for the thermal 2-point
equal-time correlation functions are also easily obtained
\begin{eqnarray}
g_{11}^{T}(x, x';t) \equiv \langle
\hat{\phi}(x)\hat{\phi}(x')\rangle_{\rm T} &=& \int [d k]
\coth \Biggl(\frac{\beta_{0} \Omega_{k}}{2}\Biggr)
\varphi^{*}_{k}\varphi_{k} e^{i  k \cdot( x -  x')}, \nonumber
\\ g_{22}^{T}(x, x';t) \equiv \langle
\hat{\pi}(x)\hat{\pi}(x')\rangle_{\rm T} &=& \int [d k]
\coth \Biggl(\frac{\beta_{0} \Omega_{k}}{2}\Biggr)
\dot{\varphi}^{*}_{k}\dot{\varphi}_{k}  e^{i  k \cdot( x - x')},
\nonumber\\ g_{12}^{T}(x, x';t) \equiv \langle \hat{\phi}(
x)\hat{\pi}(x')\rangle_{\rm T} &=& \int [d k] \coth
\Biggl(\frac{\beta_{0} \Omega_{k}}{2}\Biggr)
\dot{\varphi}^{*}_{k}\varphi_{k} e^{i k \cdot ( x - x')},
\nonumber
\\ g_{21}^{T}(x, x';t) \equiv \langle \hat{\pi}(x)
\hat{\phi}(x')\rangle_{\rm T} &=& \int [d k] \coth
\Biggl(\frac{\beta_{0} \Omega_k}{2}\Biggr)
\varphi^{*}_{k}\dot{\varphi}_{k} e^{i k \cdot(x - x')},
\end{eqnarray}
from which it follows that
\begin{eqnarray}
G_{11}^{T}(k,t) &=&  \coth
\Biggl(\frac{\beta_{0} \Omega_{k}}{2}\Biggr) \varphi^{*}_{
k}\varphi_{k}, \nonumber
\\ G_{22}^{T}(k,t) &=&  \coth
\Biggl(\frac{\beta_{0} \Omega_{k}}{2}\Biggr)
 \dot{\varphi}^{*}_{k}\dot{\varphi}_{k}, \nonumber
\\ G_{12}^{T}(k,t) &=& \coth
\Biggl(\frac{\beta_{0} \Omega_{k}}{2}\Biggr)
\dot{\varphi}^{*}_{k} \varphi_{k}, \nonumber
\\ G_{21}^{T}(k,t) &=& \coth
\Biggl(\frac{\beta_{0} \Omega_{k}}{2}\Biggr) \varphi^{*}_{k}
\dot{\varphi}_{k}. \label{eq gtherm}
\end{eqnarray}
We can define another conserved quantity for the thermal 2-point
correlation functions
\begin{eqnarray}
(\gamma^T)^{-2}(k) = G^T_{11}(k,t)G^T_{22}(k,t) -
\frac{1}{4}[G^T_{12}(k,t) + G^T_{21}(k,t)]^2  = \frac{1}{4}
\end{eqnarray}
As in the case of the nonequilibrium evolution, it is
straightforward to obtain the evolution equation for the {\it
thermal} 2-point correlation functions in the Hartree
approximation. Using Eqs. (\ref{eq gtherm}) and (\ref{eq tfh}), we
get
\begin{eqnarray}
\dot{G}_{11}^{T}(k,t) &=& G_{12}^{T}(k,t) + G_{21}^{T}(k,t),
\nonumber \\ \dot{G}_{22}^{T}(k,t) &=& - \Bigl[\omega_{k}^2 +
\frac{\lambda}{2}\phi_{c}^2(t) + \frac{\lambda}{2} g_{11}^{T} (0,
t)  \Bigr] [G_{12}^{T}(k,t) + G_{21}^{T}(k,t)] + {\cal
O}(\lambda^2), \nonumber \\ \dot{G}_{12}(k,t) &=&
G_{22}^{T}(k,t) - \Bigl[\omega_{k}^2 +
\frac{\lambda}{2}\phi_{c}^2(t) + \frac{\lambda}{2} g_{11}^{T} (0,
t) \Bigr] G_{11}^{T}(k,t) + {\cal O}(\lambda^2). \label{eq
GEQ}
\end{eqnarray}
The above set of equations together with Eq. (\ref{eq tfh})
describe self-consistently the evolution of the {\it thermal}
2-point correlation functions in the Hartree approximation. It
describes the evolution of the system initially in thermal
equilibrium at a temperature $T_0$, after interactions are turned
on at time $t=0$. They reduce to Eqs. (\ref{eq ph}) and (\ref{eq
gcor}) in the $T=0$ limit. It is also possible to obtain the set
of equations Eqs. (\ref{eq tfh})-(\ref{eq GEQ}) by taking the
expectation value of the LvN equations, Eq. (\ref{eq lvn}), with
respect to the coherent-thermal state with $H_{\rm G}$ being the
Gaussian (quadratic) part of the Hamiltonian obtained after
Hartree factorization of Eq. (\ref{eq free}), (but without
decomposing the field into a classical background and
fluctuations) using the same techniques employed in the example of
the anharmonic oscillator in Sec. III. The initial thermal
ensemble can be described adequately by $\hat{H}_{\rm G}$ and the
solution to Eq. (\ref{eq tfh}) can be obtained as
\begin{equation}
\varphi_{k} (t) = \frac{1}{\sqrt{2 \Omega_{k}}} e^{- i \Omega_{k}
t},
\end{equation}
where $\Omega_{k}$ is given by the gap equation
\begin{equation}
\Omega_{k}^2 = \omega_{k}^2 + \frac{\lambda}{2} \phi_c^2 +
\frac{\lambda}{4} \int [d k'] \frac{1}{\Omega_{k'}}\coth
\frac{\beta_0 \Omega_{k'}}{2}.
\end{equation}
For $t=0$, we then have $\varphi_{k}(0) = 1/ \sqrt{2 \Omega_k }$.
The initial thermal 2-point correlation functions are then given
by
\begin{eqnarray}
G_{11}^{T}(k,0) &=& \frac{1}{2\Omega_{k}}\Biggl[1 +
\frac{2}{e^{\beta_{0} \Omega_{k}} - 1} \Biggr], \nonumber
\\ G_{22}^{T}(k,0) &=& \frac{\Omega_{k}}{2}\Biggl[1
+ \frac{2}{e^{\beta_{0} \Omega_{k}} - 1} \Biggr], \\
G_{12}^{T}(k,0) &=& \frac{i}{2}\Biggl[1 +
\frac{2}{e^{\beta_{0} \Omega_{k}} - 1} \Biggr]. \nonumber
\end{eqnarray}
It is useful to compare our results with those obtained in
\cite{aarts,arts2}, in the context of thermalization in classical
field theories. After neglecting $\lambda$-dependent terms, the
solution of Eq. (\ref{eq tfh}) is $\varphi_{k}(t) = e^{- i
\omega_k t}/\sqrt{2\omega_k}$. With this form of $\varphi_{k}$,
one can easily show that in the classical limit ($\hbar
\rightarrow 0$) the 2-point correlation functions reduce to
\begin{eqnarray}
G_{11}^{T}(k,0) &=& \frac{T_0}{\omega_{k}^2}, \nonumber
\\ G_{22}^{T}(k,0) &=& T_{0}, \\ G_{12}^{T}(k,0) &=&
G_{21}^{T*}(k,0) = \frac{iT_{0}}{\omega_{k}}. \nonumber
\end{eqnarray}
These useful relations, which are valid for a classical thermal
system with unbroken symmetry, have been used as a benchmark to study
thermalization in classical field theory \cite{aarts,arts2}. It is possible
to carry out numerically, a spectral analysis of 2-point, equal-time, field
and momenta correlation functions and track their dynamical
evolution. The flattening out (momentum independence) of the
Fourier transform of the $\pi\pi$ equal-time correlation function
would be an indicator of thermalization in classical field theory.

\section{Nonequilibrium Evolution Beyond the Leading Order}

In this section we use two different approaches to discuss the
nonequilibrium evolution beyond the leading order Hartree
approximation. First, we use the LvN formalism to obtain expressions
for the 2-point and 4-point functions correct to ${\cal
O}(\lambda^2)$ by including all quartic terms of the Hamiltonian
and solving the time-dependent Schr\"{o}dinger equation as in Sec. IV. 
This method unifies both the LvN formalism and the Schr\"{o}dinger 
picture because the LvN formalism is only used at the Hartree approximation
and all the non-Gaussian contributions from quartic terms are found in the 
Schr\"{o}dinger picture. The second stage is similar to the interaction 
picture, though all states are expressed in terms of the time-dependent 
Hartree basis. It is found that the 2-point functions have non-Gaussian 
contribution at order of ${\cal O} (\lambda^2)$, confirming the result that 
the non-Gaussian effects appear only at ${\cal O}(\lambda^2)$
\cite{kim-khanna2}. Secondly, we make use of the Heisenberg formalism 
to obtain a set of evolution equations for the connected correlation 
functions beyond the leading order. We find that the set of equations 
form an infinite hierarchy, akin to the BBGKY hierarchy in statistical 
mechanics \cite{balescu} and incorporating next-to-leading order effects 
amounts to truncating the hierarchy at the level of the 4-point 
connected correlation functions. 

The key differences between the two formalisms lie in
the fact that while the LvN formalism (like the Schr\"{o}dinger
formalism) deals with evolution of states (with the field and
conjugate momentum operators being time-independent), the
Heisenberg formalism is associated with evolution of field and
conjugate momentum operators (with the quantum states being
time-independent). However, even in the LvN formalism, certain
operators like the annihilation and creation operators and any
function of them satisfy the LvN time-evolution equation distinguished
from the Heisenberg equation through a difference in sign. In
fact, the LvN equations can be thought of as the backward
time-evolution of the Heisenberg equations \cite{balian}.
Furthermore, the LvN approach provides a perturbative method for
going beyond the Gaussian approximation and is useful for
calculating non-Gaussian effects on domain growth in theories with
perturbatively small self-coupling constant \cite{kim-khanna2}. On the
other hand, the Heisenberg formalism provides a systematic,
non-perturbative approach for going beyond the leading order Hartree
approximation in studying the nonequilibrium evolution of quantum
fields. Inclusion of connected n-point functions in the hierarchy
of evolution equations is effectively equivalent to a loop
expansion in powers of $\hbar^{(n-1)}$ \cite{yaffe}.

\subsection{LvN Formalism: Evolution beyond the leading order}

In the LvN formalism discussed in the previous sections, the
time-dependent vacuum state is found approximately at the lowest
order as the state that is annihilated by the annihilation 
operators of all momenta in the Hartree approximation 
which satisfy the LvN equation. Going beyond the
leading order amounts to determining the proper vacuum state of
the full non-linear, interacting field theory, which can be
expressed in terms of a complete set of number states of the
Gaussian Hamiltonian at the Hartree approximation. The Gaussian
vacuum state can be improved to any desired order by including the 
perturbation part.  The n-point functions are then obtained by 
taking the expectation value of the appropriate product of field 
and momentum operators with respect to the improved vacuum state.

The normal ordered Hamiltonian is decomposed into the quadratic
Gaussian part and the quartic perturbation
\begin{equation}
\hat{H} = \hat{H}_{\rm G} + \lambda^2 \hat{H}_{\rm P},
\end{equation}
where the Gaussian part is
\begin{eqnarray}
\hat{H}_{\rm G} &=& \  \int [d k]
\Biggl[\frac{1}{2}\Biggl\{\dot{\varphi}_{-k}\dot{\varphi}_{k} +
\Biggl(\omega_{k}^2 + \frac{\lambda}{2}\int [dk_1]
\varphi^{*}_{k_1}\varphi_{k_1}\Biggr) \varphi_{-k} \varphi_{k}
\Biggr\} \hat{a}_{- k}\hat{a}_{k} \nonumber \\ && +
\frac{1}{2}\Biggl\{\dot{\varphi}_{-k}^{*}\dot{\varphi}_{k}^{*} +
\Biggl(\omega_{k}^2 + \frac{\lambda}{2}\int [dk_1]
\varphi^{*}_{k_1}\varphi_{k_1}\Biggr)
\varphi_{-k}^{*}\varphi_{k}^{*}\Biggr\}
\hat{a}_{-k}^{\dagger}\hat{a}_{k}^{\dagger} \nonumber\\&& +
\Biggl\{\dot{\varphi}^{*}_{k}\dot{\varphi}_{k} +
\Biggl(\omega_{k}^2 + \frac{\lambda}{2}\int [dk_1]
\varphi^{*}_{k_1}\varphi_{k_1}\Biggr)
\varphi_{k}^{*}\varphi_{k}\Biggr\}
\hat{a}_{k}^{\dagger}\hat{a}_{k}\Biggr]
\end{eqnarray}
and the perturbation
\begin{eqnarray}
\hat{H}_{\rm P} &=& \frac{1}{4!} \int [dk_{1}][dk_2][d k_3] [dk_4]
\delta (k_1 + k_2 + k_3 + k_4) \Bigg[ \varphi_{k_1}\varphi_{k_2}\varphi_{k_3}
\varphi_{k_4}\hat{a}_{k_1}\hat{a}_{k_2}\hat{a}_{k_3}\hat{a}_{k_4} \nonumber \\ 
&+& 4\varphi_{-k_1}^{*}\varphi_{k_2}\varphi_{k_3}\varphi_{k_4}
\hat{a}_{- k_1}^{\dagger} \hat{a}_{k_2}\hat{a}_{k_3}\hat{a}_{k_4}
+ 6\varphi_{-k_1}^{*}\varphi_{-k_2}^{*}\varphi_{k_3}\varphi_{k_3}\hat{a}_{-k_1}^{\dagger}
\hat{a}_{-k_2}^{\dagger} \hat{a}_{k_3}\hat{a}_{k_4} \nonumber\\ 
&+& 4\varphi_{- k_1}^{*}\varphi_{- k_2}^{*}\varphi^*_{-k_3}\varphi_{k_4}
\hat{a}_{- k_1}^{\dagger} \hat{a}_{-k_2}^{\dagger}\hat{a}_{- k_3}^{\dagger}
\hat{a}_{k_4} + \varphi_{- k_1}^{*}\varphi_{- k_2}^{*}\varphi_{-
k_3}^{*}\varphi_{- k_4}^{*} \hat{a}_{- k_1}^{\dagger}\hat{a}_{-
k_2}^{\dagger}\hat{a}_{- k_3}^{\dagger} \hat{a}_{- k_4}^{\dagger} \Biggr] \label{eq-ham}
\end{eqnarray}

We define the improved vacuum state as
\begin{eqnarray}
|0,t \rangle &=& \sum_{l=0}^{\infty}\sum_{n_1,n_2,...}\sum_{k_1,k_2,...}\lambda^l
C^{(l)}_{0,n_1,...}|n_1,n_2,..;t\rangle_{G} \label{vac}
%&=& \sum_{l=0}^{\infty}\sum_{n_1,n_2,...}\lambda^lC^{(l)}_{0,n_1,...}\prod_{j}
%\frac{\hat{a}_{k_j}^{\dagger n_j}}{\sqrt{n_j!}}|0;t\rangle_{\rm G} \label{vac}
\end{eqnarray}
where we have used a concise notation to represent $n_1$ particles
with momentum $k_1$, $n_2$ particles with momentum $k_2$ and so
on, for the state and the coefficient. The summation over momentum is for 
modes which have non-vanishing particle numbers.The subscript `G' refers 
to the Gaussian vacuum state.For the rest of this subsection, we will consider
the momentum space to be a discrete set and consequently, the Dirac delta functions 
will be replaced by the Kronecker delta functions. As we shall see, the coefficients 
$C^{l}_{0,n_1,...}$ which determine the vacuum state of the full non-linear theory, 
are not all non-vanishing, by requiring that the exact vacuum state
$|0,t\rangle$ satisfies the Schr\"{o}dinger equation for the full
Hamiltonian i.e.
\begin{equation}
i\frac{\partial}{\partial t} |0;t\rangle = (\hat{H}_G + \lambda \hat{H}_P)|0;t\rangle
\end{equation}
Using Eq.(\ref{vac}) we get
\begin{equation}
\sum_{l=0}^{\infty}\sum_{n_1,n_2,...}\sum_{k_1,k_2,...}
i\lambda^l\dot{C}^{(l)}_{0,n_1,...}|n_1,n_2,..;t\rangle_{G}
=  \sum_{l=0}^{\infty}\sum_{n_1,n_2,...}\sum_{k_1,k_2,...}
\lambda^{l+1}C^{(l)}_{0,n_1,...}\hat{H}_P|n_1,n_2,..;t\rangle_{G} \label{eqc1}
\end{equation}
Comparing the coefficients of $\lambda^l$ on both sides, we get
$C^{(0)}_{0;n_1,n_2,..,n_j,..}=\delta_{0,n_1}\delta_{0,n_2}....\delta_{0,n_j}....$ Then
Eq.(\ref{eqc1}) reduces to
\begin{equation}
\sum_{l=1}^{\infty}\sum_{n_1,n_2,...}\sum_{k_1,k_2,...}
i \lambda^l \dot{C}^{(l)}_{0,n_1,...}|n_1,n_2,..;t\rangle_{G}
=  \sum_{l=1}^{\infty}\sum_{n_1,n_2,...}\sum_{k_1,k_2,...}
\lambda^{l}C^{(l-1)}_{0,n_1,...}\hat{H}_P|n_1,n_2,..;t\rangle_{G}
\end{equation}
Equating the coefficients of the same power of $\lambda$ on both
sides, we get the following equation for the expansion
coefficients
\begin{equation}
\dot{C}^{(l)}_{0;n_1,n_2,...}(t) = -i \sum_{m_{1^{\prime}},...,m_{j^{\prime}},...}
\sum_{k_1^{\prime},...,k_j^{\prime},...}
C^{(l-1)}_{0;m_{1^{\prime}},...m_{j^{\prime}},..}(t)
{}_{\rm G}\langle n_1...n_j...;t|\hat{H}_P|m_{1^{\prime}}...m_{j^{\prime}}...;t \rangle_{G} 
\label{eq-c2}
\end{equation}
The vacuum state correct to ${\cal O}(\lambda^2)$ then becomes
\begin{eqnarray}
|0;t\rangle_{[2]} &=& |0;t\rangle_{G} + \lambda\sum_{n_1,n_2,...n_j,...}\sum_{k_1,k_2,...k_j,...}
C^{(1)}_{0;n_1,..,n_j,..}(t)|n_1,..,n_j,..;t\rangle_G \nonumber \\ &+& 
\lambda^2\sum_{n_1,n_2,...n_j,...}\sum_{k_1,k_2,...k_j,...}C^{(2)}_{0;n_1,..,n_j,..}(t)
|n_1,..,n_j,..;t\rangle_G \label{vac2}
\end{eqnarray}
From Eq.(\ref{eq-ham}) and Eq.(\ref{eq-c2}), the only non-vanishing
contribution to $C^{(1)}_{0;n_1,..,n_j,..}(t)$ comes from the
following term $\frac{1}{4!}\int [dk_1][dk_2][dk_3][dk_4]\delta(k_1+k_2+k_3+k_4)
\varphi^*_{-k_1}\varphi^*_{-k_2}\varphi^*_{-k_3}\varphi^*_{-k_4}
\hat{a}_{-k_1}^{\dagger}\hat{a}_{-k_2}^{\dagger}\hat{a}_{-k_3}^{\dagger}\hat{a}_{-k_4}^{\dagger}$,
where the indices correspond to excitations in modes
$k_1,k_2,k_3$.  
%To obtain the coefficients $C^{(1)}$ and $C^{(2)}$ we consider two different
%relations between the modes $k_1,k_2,k_3$. These are
%(i)$k_3=-k_2=k_1$ and (ii)$k_3 = -k_2 \neq k_1$.
The most general form of the equation for the first order coefficient follows from 
Eq.(\ref{eq-c2}). It can be written as 
\begin{eqnarray}
\dot{C}^{(1)}_{0;\{n_{k}\}}=-\frac{i}{4!}(N_f)(N_c)
\sum_{k_1^{\prime},k_2^{\prime}..}\langle n_1,..,n_j,..;t|
\sum_{k_1^{\prime\prime},k_2^{\prime\prime}..}
\delta_{k_1^{\prime\prime}+k_2^{\prime\prime}+k_3^{\prime\prime}+k_4^{\prime\prime},0}
\varphi^*_{k_1^{\prime\prime}}\varphi^*_{k_2^{\prime\prime}}\varphi^*_{k_3^{\prime\prime}}
\varphi^*_{k_4^{\prime\prime}}
\hat{a}_{k_1^{\prime\prime}}^{\dagger}\hat{a}_{k_2^{\prime\prime}}^{\dagger}
\hat{a}_{k_3^{\prime\prime}}^{\dagger}\hat{a}_{k_4^{\prime\prime}}^{\dagger}
|0;t\rangle     
\label{eqcf1}
\end{eqnarray}
where $\{n_k\}\equiv n_{k_1},n_{k_2},n_{k_3},n_{k_4}$ and $N_c$ and $N_f$ are numerical 
coefficients which depend on the relationship between the modes $k_1,k_2,k_3,k_4$. 
$N_c$ correspond to the factor determined by the maximum power of a creation operator, 
whereas $N_f$ is a combinatorial factor. It is important to note that {\it distinct} 
first-order coefficients (with distinct values of $n_{k_1},n_{k_2}$ etc.) arise when 
constraints are imposed on the momentum modes $k_1,k_2,k_3,k_4$ which account for 
different types of scattering processes. In the absence of any constraints, 
$N_{c}=N_f=1$ and Eq.(\ref{eqcf1}) gives 
\begin{equation}
\dot{C}^{(1)}_{0;1_{k_1},1_{k_2},1_{k_3},1_{k_4}}=-\frac{i}{4!}\varphi^*_{k_1}
\varphi^*_{k_2}\varphi^*_{k_3}\varphi^*_{k_4}\delta_{k_1+k_2+k_3+k_4,0}
\end{equation}
When the four modes appearing as indices in the creation operators are paired resulting in two 
distinct creation operators (i.e. when $k_2^{\prime\prime}=k_1^{\prime\prime}$ and 
$k_4^{\prime\prime}=k_3^{\prime\prime}$), $N_c=(\sqrt{2})^2$ and $N_f=(4_{C_2}/2)\times2_{C_2}$
(since there are 3 ways of forming 2 distinct pairs of creation operators); 
Eq.(\ref{eqcf1}) becomes  
\begin{equation}
\dot{C}^{(1)}_{0;2_{k_1},2_{k_2}} = -\frac{3i}{4!}(\sqrt{2})^2
(\varphi_{k_1}^*\varphi_{k_2}^*)^2\delta_{k_1+k_2,0}
\end{equation}
For a single pairing (i.e. when for example $k_2^{\prime\prime}=k_1^{\prime\prime}$ but 
$k_4^{\prime\prime}\neq k_3^{\prime\prime}$), $N_c=\sqrt{2}$ and $N_f=4_{C_2}$ 
(since there are $4_{C_2}$ ways of forming a single pair out of 4 creation operators) and 
Eq.(\ref{eqcf1}) becomes  
\begin{equation}
\dot{C}^{(1)}_{0;2_{k_1},1_{k_2},1_{k_3}} = -\frac{6i}{4!}\sqrt{2}(\varphi_{k_1}^{*2})
\varphi_{k_2}^*\varphi_{k_3}^*\delta_{2k_1+k_2+k_3,0}
\end{equation}
From the form of the string of creation operators appearing in Eq.(\ref{eqcf1}), it is 
clear that the number of different coefficients correspond to the different ways of 
partitioning of four particle excitations in terms of excitations of lower order.  

In order to derive non-Gaussian corrections to the 2-point functions, it is necessary to 
obtain the corrected vacuum state, at least upto ${\cal O}(\lambda^2)$. For this purpose, 
we need to determine the second order coefficients. Although, there are many 
distinct second order coefficients, we will see that only a few of them will contribute 
to the 2-point functions, thereby considerably simplifying the calculations.Nevertheless,
for the sake of completeness, we explicitly outline below, the general method for 
obtaining all possible second order coefficients. The equation for the second order 
coefficient can be written in the most general form 
\begin{eqnarray}
\dot{C}^{(2)}_{0;\{n_{k}\}}=-i \sum_{k_{1}^\prime,k_2^{\prime},..}
C^{(1)}_{0;\{m_{k^{\prime}}\}}
\delta_{k_1^{\prime}+k_2^{\prime}+k_3^{\prime}+k_4^{\prime},0}
\langle n_1,..,n_j,..;t|\hat{H}_P\hat{a}_{k_1^{\prime}}^{\dagger}
\hat{a}_{k_2^{\prime}}^{\dagger}\hat{a}_{k_3^{\prime}}^{\dagger}
\hat{a}_{k_4^{\prime}}^{\dagger}|0;t \rangle
\label{eqcf2}
\end{eqnarray}
It is clear from the string of annihilation and creation operators appearing
in each term in Eq.(\ref{eq-ham}) that the second order coefficients can be 
divided into five classes corresponding to zero,two,four,six and eight particle 
excitations respectively. The zero-particle excitation case leads to just 
one second order coefficient given by the equation 
\begin{eqnarray}
\dot{C}^{(2)}_{0;0} &=& -\frac{i}{4!}\sum_{k_{1}^\prime,k_2^{\prime},..}
C^{(1)}_{0;1_{k_1^{\prime}},1_{k_2^{\prime}},1_{k_3^{\prime}},1_{k_4^{\prime}}}
\delta_{k_1^{\prime}+k_2^{\prime}+k_3^{\prime}+k_4^{\prime},0} \nonumber \\
&&\langle n_1,..,n_j,..;t|\sum_{k_{1}^{\prime\prime},k_2^{\prime\prime},..}
\delta_{k_1^{\prime\prime}+k_2^{\prime\prime}+k_3^{\prime\prime}+k_4^{\prime\prime},0}
\varphi_{k_1^{\prime\prime}}\varphi_{k_2^{\prime\prime}}\varphi_{k_3^{\prime\prime}}
\varphi_{k_4^{\prime\prime}}
\hat{a}_{k_1^{\prime\prime}}\hat{a}_{k_2^{\prime\prime}}
\hat{a}_{k_3^{\prime\prime}}\hat{a}_{k_4^{\prime\prime}}
\hat{a}_{k_1^{\prime}}^{\dagger}\hat{a}_{k_2^{\prime}}^{\dagger}
\hat{a}_{k_3^{\prime}}^{\dagger}\hat{a}_{k_4^{\prime}}^{\dagger}|0;t\rangle 
\end{eqnarray}
By making use of the commutation relations between the annihilation and creation 
operators to normal order the above string of operators,it can be easily shown 
that the above equation reduces to
\begin{equation}
\dot{C}^{(2)}_{0;0}=-iC^{(1)}_{0;1_{k_1},1_{k_2},1_{k_3},1_{k_4}}
\delta_{k_1+k_2+k_3+k_4,0}\varphi_{k_1}\varphi_{k_2}\varphi_{k_3}\varphi_{k_4}
\end{equation}

 The second order coefficients corresponding to the two particle excitation case, 
which is contributed by the second term in Eq.(\ref{eq-ham}) can be written in general 
as

\begin{eqnarray}
\dot{C}^{(2)}_{\{2_k\};0} &=& -4i\sum_{k_{1}^\prime,k_2^{\prime},..}
C^{(1)}_{0;1_{k_1^{\prime}},1_{k_2^{\prime}},1_{k_3^{\prime}},1_{k_4^{\prime}}}
\delta_{k_1^{\prime}+k_2^{\prime}+k_3^{\prime}+k_4^{\prime},0} \nonumber \\
&&\langle n_1,..,n_j,..;t|\sum_{k_{1}^{\prime\prime},k_2^{\prime\prime},..}
\delta_{-k_1^{\prime\prime}+k_2^{\prime\prime}+k_3^{\prime\prime}+k_4^{\prime\prime},0}
\delta_{k_2^{\prime},k_2^{\prime\prime}}\delta_{k_3^{\prime},k_3^{\prime\prime}}
\delta_{k_4^{\prime},k_4^{\prime\prime}}
\varphi^*_{k_1^{\prime\prime}}\varphi_{k_2^{\prime\prime}}\varphi_{k_3^{\prime\prime}}
\varphi_{k_4^{\prime\prime}}
\hat{a}^{\dagger}_{k_1^{\prime\prime}}\hat{a}^{\dagger}_{k_1^{\prime}}|0;t\rangle
\end{eqnarray} 
which leads to the equations for the following second-order coefficient
\begin{eqnarray}
%\dot{C}^{(2)}_{0;2_{k_1}} &=& -4i\sqrt{2}C^{(1)}_{0;1_{k_1},1_{k_2},1_{k_3},1_{k_4}}
%\delta_{-k_1+k_2+k_3+k_4,0}\varphi^{*}_{k_1}\varphi_{k_2}\varphi_{k_3}\varphi_{k_4} 
%\nonumber \\
\dot{C}^{(2)}_{0;1_{k_1},1_{k_5}} &=& -4iC^{(1)}_{0;1_{k_1},1_{k_2},1_{k_3},1_{k_4}}
\delta_{k_1+k_2+k_3+k_4,0}\delta_{k_2+k_3+k_4-k_5,0}\varphi^{*}_{k_5}
\varphi_{k_2}\varphi_{k_3}\varphi_{k_4}
\end{eqnarray}

 The third term in Eq.(\ref{eq-ham}) leads to the following general equation for
the second-order coefficients associated with the four particle excitation case
\begin{eqnarray}
\dot{C}^{(2)}_{\{4_k\};0} &=& -3i\sum_{k_{1}^\prime,k_2^{\prime},..}
C^{(1)}_{0;1_{k_1^{\prime}},1_{k_2^{\prime}},1_{k_3^{\prime}},1_{k_4^{\prime}}}
\delta_{k_1^{\prime}+k_2^{\prime}+k_3^{\prime}+k_4^{\prime},0} \nonumber \\
&&\langle n_1,..,n_j,..;t|\sum_{k_{1}^{\prime\prime},k_2^{\prime\prime},..}
\delta_{-k_1^{\prime\prime}-k_2^{\prime\prime}+k_3^{\prime\prime}+k_4^{\prime\prime},0}
\delta_{k_3^{\prime},k_3^{\prime\prime}}\delta_{k_4^{\prime},k_4^{\prime\prime}}
\varphi_{k_1^{\prime\prime}}^*\varphi_{k_2^{\prime\prime}}^*\varphi_{k_3^{\prime\prime}}
\varphi_{k_4^{\prime\prime}}
\hat{a}^{\dagger}_{k_1^{\prime\prime}}\hat{a}^{\dagger}_{k_2^{\prime\prime}}
\hat{a}^{\dagger}_{k_1^{\prime}}\hat{a}^{\dagger}_{k_2^{\prime}}|0;t\rangle
\end{eqnarray} 
which leads to the dynamical equations for the following four second-order coefficients
\begin{eqnarray}
\dot{C}^{(2)}_{0;2_{k_1},2_{k_2}} &=& -9i(\sqrt{2})^2
C^{(1)}_{0;1_{k_1},1_{k_2},1_{k_3},1_{k_4}}
\delta_{k_1+k_2+k_3+k_4,0}\delta_{-k_1+k_2+k_3+k_4,0}
\varphi^{*}_{k_1}\varphi^*_{k_2}\varphi_{k_3}\varphi_{k_4} \nonumber \\
\dot{C}^{(2)}_{0;2_{k_1},1_{k_2},1_{k_5}} &=& -18i\sqrt{2}
C^{(1)}_{0;1_{k_1},1_{k_2},1_{k_3},1_{k_4}}
\delta_{k_1+k_2+k_3+k_4,0}\delta_{-k_1+k_3+k_4-k_5,0}
\varphi^{*}_{k_1}\varphi^*_{k_5}\varphi_{k_3}\varphi_{k_4} \nonumber \\
\dot{C}^{(2)}_{0;1_{k_1},1_{k_2},1_{k_5},1_{k_6}} &=& -3i
C^{(1)}_{0;1_{k_1},1_{k_2},1_{k_3},1_{k_4}}
\delta_{k_1+k_2+k_3+k_4,0}\delta_{k_3+k_4-k_5-k_6,0}
\varphi^{*}_{k_5}\varphi^*_{k_6}\varphi_{k_3}\varphi_{k_4} \nonumber \\
\dot{C}^{(2)}_{0;3_{k_1},1_{k_2}} &=& -12i(\sqrt{6})
C^{(1)}_{0;1_{k_1},1_{k_2},1_{k_3},1_{k_4}}
\delta_{-2k_1+k_3+k_4,0}\delta_{k_1+k_2+k_3+k_4,0}
\varphi^{*2}_{k_1}\varphi_{k_3}\varphi_{k_4}
\end{eqnarray}

 Similarly, the second order coefficients corresponding to the six and eight particle 
excitations contributed by the last two terms in  Eq.(\ref{eq-ham}) respectively can
be easily obtained. The general structure of the second order coefficients 
corresponding to the six particle excitation case is given by 

\begin{eqnarray}
\dot{C}^{(2)}_{\{6_k\};0} &=& -\frac{2i}{3}\sum_{k_{1}^\prime,k_2^{\prime},..}
C^{(1)}_{0;1_{k_1^{\prime}},1_{k_2^{\prime}},1_{k_3^{\prime}},1_{k_4^{\prime}}}
\delta_{k_1^{\prime}+k_2^{\prime}+k_3^{\prime}+k_4^{\prime},0} \nonumber \\
&&\langle n_1,..,n_j,..;t|\sum_{k_{1}^{\prime\prime},k_2^{\prime\prime},..}
\delta_{-k_1^{\prime\prime}-k_2^{\prime\prime}-k_3^{\prime\prime}+k_4^{\prime\prime},0}
\delta_{k_4^{\prime},k_4^{\prime\prime}}
\varphi_{k_1^{\prime\prime}}^*\varphi_{k_2^{\prime\prime}}^*
\varphi_{k_3^{\prime\prime}}^*\varphi_{k_4^{\prime\prime}}
\hat{a}^{\dagger}_{k_1^{\prime\prime}}\hat{a}^{\dagger}_{k_2^{\prime\prime}}
\hat{a}^{\dagger}_{k_3^{\prime\prime}}
\hat{a}^{\dagger}_{k_1^{\prime}}\hat{a}^{\dagger}_{k_2^{\prime}}
\hat{a}^{\dagger}_{k_3^{\prime}}|0;t\rangle
\end{eqnarray}
and that associated with the eight particle excitation case is given by
\begin{eqnarray}
\dot{C}^{(2)}_{\{8_k\};0} &=& -\frac{i}{4!}\sum_{k_{1}^\prime,k_2^{\prime},..}
C^{(1)}_{0;1_{k_1^{\prime}},1_{k_2^{\prime}},1_{k_3^{\prime}},1_{k_4^{\prime}}}
\delta_{k_1^{\prime}+k_2^{\prime}+k_3^{\prime}+k_4^{\prime},0} \nonumber \\
&&\langle n_1,..,n_j,..;t|\sum_{k_{1}^{\prime\prime},k_2^{\prime\prime},..}
\delta_{k_1^{\prime\prime}+k_2^{\prime\prime}+k_3^{\prime\prime}+k_4^{\prime\prime},0}
\varphi_{k_1^{\prime\prime}}^*\varphi_{k_2^{\prime\prime}}^*
\varphi_{k_3^{\prime\prime}}^*\varphi_{k_4^{\prime\prime}}^*
\hat{a}^{\dagger}_{k_1^{\prime\prime}}\hat{a}^{\dagger}_{k_2^{\prime\prime}}
\hat{a}^{\dagger}_{k_3^{\prime\prime}}\hat{a}^{\dagger}_{k_4^{\prime\prime}}
\hat{a}^{\dagger}_{k_1^{\prime}}\hat{a}^{\dagger}_{k_2^{\prime}}
\hat{a}^{\dagger}_{k_3^{\prime}}\hat{a}^{\dagger}_{k_4^{\prime}}|0;t\rangle
\end{eqnarray}
However, none of those coefficients contribute to the expressions
for the two-point functions correct to ${\cal O}(\lambda^2)$. Hence, we do not list
them explicitly. 

 The normalization of the vacuum state correct to second order in $\lambda$ is 
easily found in terms of the first and second order coefficients
\begin{eqnarray}
{}_{[2]}\langle 0;t|0;t \rangle_{[2]} &=& 1+ \lambda^2 \biggr[\biggl\{ 
\sum_{k_1,k_2}C^{(1)*}_{0;2_{k_1},2_{k_2}}C^{(1)}_{0;2_{k_1},2_{k_2}} 
+ \sum_{k_1,k_2,k_3}C^{(1)*}_{0;2_{k_1},1_{k_2},1_{k_3}}
C^{(1)}_{0;2_{k_1},1_{k_2},1_{k_3}} \nonumber \\ &+& \sum_{k_1,k_2,k_3,k_4}
C^{(1)*}_{0;1_{k_1},1_{k_2},1_{k_3},1_{k_4}}
C^{(1)}_{0;1_{k_1},1_{k_2},1_{k_3}1_{k_4}}\biggr\} 
+ \biggl\{ C^{(2)*}_{0;0}+C^{(2)}_{0;0} \biggr\}
\biggr] + {\cal O}(\lambda^3) 
\end{eqnarray}

 We can now obtain expressions for the equal-time, 2-point
functions by taking the expectation values with respect to this
improved, albeit perturbative, vacuum state, with the understanding
that
\begin{equation}
G^{(2)}_{ab}(k) = {}_{\rm [2]}\langle0;t|\hat{\Phi}_a(k)
\hat{\Phi}_b(-k)|0;t\rangle_{\rm [2]}
\end{equation}
where the subscript indices $a,b, \cdots =1,2$ and
$\hat{\Phi}_{a}$ is to be understood as the field operator
$\hat{\Phi}(k)$ ($a=1$) or the conjugate momentum operator 
$\hat{\Pi}(k)$($a=2$). Moreover we will be working with the symmetric 
theory which implies that $\phi_c=0$. It is important to note that in the LvN 
formalism, the field and momentum operators are {\it time-independent}. The 
time dependence of the 2-point functions arise through the time-dependence of the
auxiliary field variable $\varphi$. It is then easy to show that the
equal-time, 2-point functions are given up to and including ${\cal
O}(\lambda^2)$ by the equations

\begin{eqnarray}
G^{(2)}_{11}(k;t) &=& \varphi_{k}^{*}\varphi_{k}\biggl[1 +
\lambda^{2}\biggl\{\sum_{k_1,k_2}C^{(1)*}_{0;2_{k_1},2_{k_2}}C^{(1)}_{0;2_{k_1},2_{k_2}} 
+ \sum_{k_1,k_2,k_3}C^{(1)*}_{0;2_{k_1},1_{k_2},1_{k_3}}C^{(1)}_{0;2_{k_1},1_{k_2},1_{k_3}} \nonumber \\ 
&+& \sum_{k_1,k_2,k_3,k_4}
C^{(1)*}_{0;1_{k_1},1_{k_2},1_{k_3},1_{k_4}}
C^{(1)}_{0;1_{k_1},1_{k_2},1_{k_3},1_{k_4}} + 
(C^{(2)*}_{0;0} + C^{(2)}_{0;0})\biggr\}\biggr] 
+ \lambda^2\biggl[2C^{(2)}_{0;1_{-k},1_k}\varphi_{-k}\varphi_{k} \nonumber \\
&+& 4\sum_{k_1}\biggl(C^{(1)*}_{0;2_{-k},2_{k_1}}C^{(1)}_{0;2_{-k},2_{k_1}}
\varphi^{*}_{-k}\varphi_{-k} + C^{(1)*}_{0;2_{k},2_{k_1}}C^{(1)}_{0;2_{k},2_{k_1}}
\varphi^{*}_{k}\varphi_{k}\biggr)\biggr] 
+ {\cal O}(\lambda^3)
\end{eqnarray}
\begin{eqnarray}
G^{(2)}_{21}(k;t) &=& \varphi_{k}^{*}\dot{\varphi}_{k}\biggl[1 +
\lambda^{2}\biggl\{\sum_{k_1,k_2}C^{(1)*}_{0;2_{k_1},2_{k_2}}C^{(1)}_{0;2_{k_1},2_{k_2}} 
+ \sum_{k_1,k_2,k_3}C^{(1)*}_{0;2_{k_1},1_{k_2},1_{k_3}}C^{(1)}_{0;2_{k_1},1_{k_2},1_{k_3}} \nonumber \\ 
&+& \sum_{k_1,k_2,k_3,k_4}
C^{(1)*}_{0;1_{k_1},1_{k_2},1_{k_3},1_{k_4}}
C^{(1)}_{0;1_{k_1},1_{k_2},1_{k_3},1_{k_4}} + 
(C^{(2)*}_{0;0} + C^{(2)}_{0;0})\biggr\}\biggr] 
+ \lambda^2\biggl[2C^{(2)}_{0;1_{-k},1_k}\varphi_{-k}\dot{\varphi}_{k} \nonumber \\
&+& 4\sum_{k_1}\biggl(C^{(1)*}_{0;2_{-k},2_{k_1}}C^{(1)}_{0;2_{-k},2_{k_1}}
\dot{\varphi}^{*}_{-k}\varphi_{-k} + C^{(1)*}_{0;2_{k},2_{k_1}}C^{(1)}_{0;2_{k},2_{k_1}}
\varphi^{*}_{k}\dot{\varphi}_{k}\biggr)\biggr]
+ {\cal O}(\lambda^3) 
\end{eqnarray}
\begin{eqnarray}
G^{(2)}_{22}(k;t) &=& \dot{\varphi}_{k}^{*}\dot{\varphi}_{k}\biggl[1 +
\lambda^{2}\biggl\{\sum_{k_1,k_2}C^{(1)*}_{0;2_{k_1},2_{k_2}}C^{(1)}_{0;2_{k_1},2_{k_2}} 
+ \sum_{k_1,k_2,k_3}C^{(1)*}_{0;2_{k_1},1_{k_2},1_{k_3}}C^{(1)}_{0;2_{k_1},1_{k_2},1_{k_3}} \nonumber \\ 
&+& \sum_{k_1,k_2,k_3,k_4}
C^{(1)*}_{0;1_{k_1},1_{k_2},1_{k_3},1_{k_4}}
C^{(1)}_{0;1_{k_1},1_{k_2},1_{k_3},1_{k_4}} + 
(C^{(2)*}_{0;0} + C^{(2)}_{0;0})\biggr\}\biggr] 
+ \lambda^2\biggl[2C^{(2)}_{0;1_{-k},1_k}\dot{\varphi}_{-k}\dot{\varphi}_{k} \nonumber \\
&+& 4\sum_{k_1}\biggl(C^{(1)*}_{0;2_{-k},2_{k_1}}C^{(1)}_{0;2_{-k},2_{k_1}}
\dot{\varphi}^{*}_{-k}\dot{\varphi}_{-k} 
+ C^{(1)*}_{0;2_{k},2_{k_1}}C^{(1)}_{0;2_{k},2_{k_1}}
\dot{\varphi}^{*}_{k}\dot{\varphi}_{k}\biggr)\biggr]
+ {\cal O}(\lambda^3)
\end{eqnarray}
It is clearly evident from this set of equations that the first
term corresponds to the Hartree approximation result and
corrections to the Hartree approximation appear only at ${\cal O}(\lambda^2)$. 
The auxiliary field mode variable $\varphi_{k}$ is given 
by the solution of Eq.(\ref{eq ph}) with $\lambda=0$. The effect of 
scattering enters through the presence of the first and second order 
coefficients which first appear at ${\cal O}(\lambda^2)$. For example, 4-particle 
scattering processes with overall momentum conservation are encoded in  
terms involving the first order coefficients 
$C^{(1)}_{0;1_{k_1},1_{k_2},1_{k_3},1_{k_4}}$. The LvN formalism thus provides a 
systematic perturbative method for computing non-Gaussian (beyond Hartree) 
corrections to the 2-point function and higher n-point functions. Systematic 
corrections in powers of $\lambda \ge 3$ can also be derived in a similar manner
after including terms of ${\cal O}(\lambda^n)$ ($n \ge 3$) in the expression
for the improved vacuum state. These expressions for the 2-point functions particularly 
suited to obtain non-Gaussian corrections to the domain size in systems undergoing 
a quenched second-order phase transition. The 4-point function and higher n-point 
functions in momentum space 
%defined by the equation
%\begin{equation}
%G^{(4)}_{abcd}(k_1^{\prime},k_2^{\prime},k_3^{\prime}) = {}_{\rm [2]}\langle0,t|
%\hat{\Phi}_{a}(k_1^{\prime})\hat{\Phi}_{b}(k_2^{\prime})\hat{\Phi}_{c}(k_3^{\prime})
%\hat{\Phi}_{d}(-k_1^{\prime}-k_2^{\prime}-k_3^{\prime})|
%0;t\rangle_{\rm [2]}
%\end{equation}
%($a,b,c,d=1,2$) 
can be also be computed in a similar manner, in terms of the auxiliary field
modes and the expansion coefficients up to ${\cal O}(\lambda^2)$.
The $\lambda$-independent terms in the 4-point functions correspond to 
the factorization of the 4-point function in terms of products of 2-point 
functions and thereby represent the Hartree approximation result.

\subsection{Heisenberg Formalism : Evolution beyond the leading order}

In this subsection, we obtain the nonequilibrium evolution
equations for the connected correlators, by taking the vacuum
expectation value of the Heisenberg equations of motion for
various combinations of products of field and its conjugate
momentum operators, after making use of the cluster expansion to
express the ordinary n-point functions as sum of products of
connected n-point functions of lower order. This technique was
used to obtain the effective potential and investigate phase
transitions in spontaneously broken $\phi^4$-theory in $1+1$ and
$2+1$ dimensions \cite{peter}. We note from Sec. II that the
expectation value of a functional of Heisenberg-operators with
respect to the vacuum state becomes
\begin{equation}
\langle 0 \vert \hat{F}_{\rm H} (t) \vert 0 \rangle = \langle 0, t \vert
\hat{F}_{\rm S} \vert 0, t \rangle,
\end{equation}
where $\hat{F}_{\rm H}(t)$ is any general function(al) of operators in the
Heisenberg picture and $\hat{F}_{\rm S}$ is the corresponding time-independent
operator in the Schr\"{o}dinger picture.

From now on, to take into account NLO effects and compare with
other methods, we adopt an approach different from the LvN
formalism. We take the expectation value of the Heisenberg
equation for any functional of $\hat{\phi}_{{\rm H}}(t)$ and
$\hat{\pi}_{{\rm H}}(t)$ as follows
\begin{equation}
\langle 0 \vert i \  \frac{\partial}{\partial t} \hat{F}_{\rm H} (t)
\vert 0 \rangle =
\langle 0 \vert [\hat{F}_{\rm H}(t),\hat{H}_{\rm H} (t)]\vert 0 \rangle
\end{equation}
where $\hat{H}_{\rm H}(t)$ and $\hat{F}_{\rm H}(t)$ are
time-dependent operators  and $\vert 0 \rangle$ is the
time-independent vacuum state. Even though the explicit form of
$\hat{H}_{\rm H}(t)$ and $\hat{F}_{\rm H}(t)$ in terms of the
Schr\"{o}dinger operators requires knowledge of the unitary
evolution operator, we will not be required to know their explicit
forms for the purpose of the calculation below.

In general, the evolution equations for the ordinary n-point
correlators which are defined as
\begin{eqnarray}
g^{(n)}_{a, \cdots, b}({\bf x}_{1}, \cdots , {\bf x}_{n};t) &=&
\langle \hat{\phi}_{a}({\bf
x_1},t) \cdots \hat{\phi}_{b}({\bf x}_n,t) \rangle \\
&\equiv&
\langle \hat{F}_{{\rm H};a, \cdots, b}({\bf x}_{1}, \cdots, {\bf x}_{n};t)
\rangle
\nonumber
\end{eqnarray}
are obtained by taking the vacuum expectation value of the
Heisenberg equation (after dropping the subscript `H')
\begin{equation}
\frac{d \langle \hat{F}_{a, \cdots, b} \rangle}{dt} = \frac{1}{i}
\langle [\hat{F}_{a, \cdots, b}({\bf x}_{1}, \cdots, {\bf
x}_{n};t),\hat{H}] \rangle \label{eveq1}
\end{equation}
where the subscript indices $a,b, \cdots =1,2$ for our model
Hamiltonian and $\hat{\phi}_{a}$ is to be understood as the
fluctuation field operator $\hat{\phi}_{f}({\bf x}, t)$ ($a=1$) or
the conjugate momentum operator $\hat{\pi}_{f}({\bf x}, t)$
($a=2$) in the Heisenberg picture. The subscript $f$ has been
removed for notational convenience. The fluctuation field and its
conjugate momenta satisfy the usual commutation relations
\begin{equation}
[\hat{\phi}_{a}({\bf x}, t),\hat{\phi}_{b}({\bf y}, t)] = i
\delta_{ab} \delta^{D}({\bf x} - {\bf y}). \label{hseq}
\end{equation}

For the unbroken symmetry case, $\phi_c=0$, only correlation
functions of even order are non-vanishing. (For a theory
exhibiting spontaneous symmetry breaking, a non-vanishing vacuum
expectation value will induce a cubic interaction term because of
which even {\it odd} n-point functions become non-trivial). Using
the cluster expansion \cite{peter} to express ordinary n-point
functions in terms of the equal-time, connected n-point
correlators, it is possible to obtain a set of evolution equations
for the connected correlators. Because of the presence of the
quartic coupling, the equations for the 2-point functions would
depend upon the connected 4-point functions; the equations for the
4-point functions would depend upon the 6-point functions and so
on, thereby yielding an infinite hierarchy of evolutions equations
for the n-point correlators. To go beyond the Hartree
approximation requires including the cluster expansion of the
6-point function in terms of products of connected n-point
functions of lower order and ignoring the effect of the connected
6-point function in the expansion. This amounts to taking into
account direct scattering effects and provides a systematic way of going
beyond the leading-order mean field expansion. This then yields a
set of closed equations which have been truncated at the 4-point
level. The leading order (Hartree approximation) result is easily
recovered by expressing the 4-point function as products of
connected 2-point functions and ignoring the term involving the
connected 4-point function in the cluster expansion. This yields a
closed set of equations for the equal-time, connected 2-point
functions. In this scheme, incorporating NNLO effects would then
amount to truncating the hierarchy of evolution equations at the
6-point level. Appendix A lists the cluster expansion of n-point
functions ($n \leq 4$) and the set of evolution equations for the
equal-time, connected n-point functions ($n \leq 4$) in
configuration space. It is often more convenient to work with the
Fourier transforms of the evolution equations. For that purpose,
we define the Fourier transforms of the 2 and 4-point functions as
\begin{eqnarray}
g^{(2C)}_{ab}(x_1,x_2) &=& \int [dk_1][dk_2] \tilde{G}^{(2C)}_{ab}(k_1,k_2)
e^{i( k_1 \cdot x_1 + k_2 \cdot x_2)} \nonumber \\
g^{(4C)}_{abcd}(x_1,x_2,x_3,x_4) &=& \int [dk_1][dk_2][dk_3][dk_4]
\tilde{G}^{(4C)}_{abcd}(k_1,k_2,k_3,k_4) e^{i(\sum_{i=1}^{i=4}
k_{i} \cdot x_{i})}
\end{eqnarray}
%For brevity of notation, we used the representation $dk \equiv
%\frac{d^{D}{\bf k}}{(2\pi)^D}$ and $x = {\bf x}$. 
For translationally invariant theories,
\begin{eqnarray}
\tilde{G}^{(2C)}_{ab}(k_1,k_2)&=&G^{(2C)}_{ab}(k_1)\delta(k_1+k_2) \nonumber\\
\tilde{G}^{(4C)}_{abcd}(k_1,k_2,k_3,k_4) &=& G^{(4C)}_{abcd}(k_1,k_2,k_3)
\delta(k_1+k_2+k_3+k_4)
\end{eqnarray}
which is just an indication of the conservation of momentum at the
vertices. The Fourier transform of Eq. (A1) and Eqs. (A3)-(A7)
yields the following evolution equations for $G^{(2C)}_{ab}(k)$:
\begin{eqnarray}
\dot{G}^{(2C)}_{11}(k_1) &=& G^{(2C)}_{12}(k_1) +
G^{(2C)}_{21}(k_1),
\nonumber \\
\dot{G}^{(2C)}_{21}(k_1) &=& G^{(2C)}_{22}(k_1) - \biggl[
\omega^2(k_1) + \frac{\lambda}{2}\int{[dk^{\prime}]
G^{(2C)}_{11}(k^{\prime}) \biggr]
G^{(2C)}_{11}(k_1)}
- \frac{\lambda}{6}\int{[dk^{\prime}][dk^{\prime\prime}]
G^{(4C)}_{1111}(k^{\prime},k^{\prime\prime},-k_1)}, \nonumber \\
\nonumber \\
\dot{G}^{(2C)}_{22}(k_1) &=& - \biggl[ \omega^2(k_1)  + \frac{\lambda}{2}
\int{[dk^{\prime}] G^{(2C)}_{11}(k^{\prime}) \biggr] \biggl( G^{(2C)}_{12}(k_1) +
G^{(2C)}_{21}(k_1)\biggr)} \nonumber \\
&& - \frac{\lambda}{6}\int{[dk^{\prime}][dk^{\prime\prime}] \biggl(
G^{(4C)}_{1112}(k^{\prime},k^{\prime\prime},k_1) +
G^{(4C)}_{2111}(k^{\prime},k^{\prime\prime},k_1) \biggr)}.
\end{eqnarray}

By neglecting terms involving the connected 4-point functions
in the above equations and setting $\phi_c=0$ (since we are discussing the
symmetric theory), one recovers the leading order (Hartree approximation)
evolution equations, Eq.\ref{eq gcor}, obtained in the last section. This also
establishes the equivalence of the method used in this section to obtain
NLO equations and the LvN method used in the previous sections.
The evolution equations for the 4-point functions
$G^{(4C)}_{abcd}(k_1,k_2,k_3)$ obtained by taking the Fourier
transform of Eqs. (A3)-(A7) are
\begin{eqnarray}
\dot{G}^{(4C)}_{1111}(k_1,k_2,k_3) &=&
G^{(4C)}_{2111}(k_1,k_2,k_3) + G^{(4C)}_{1211}(k_1,k_2,k_3) +
G^{(4C)}_{1121}(k_1,k_2,k_3) + G^{(4C)}_{1112}(k_1,k_2,k_3), \\
\dot{G}^{(4C)}_{2111}(k_1,k_2,k_3) &=& G^{(4C)}_{2211}(k_1,k_2,k_3) +
G^{(4C)}_{2121}(k_1,k_2,k_3) + G^{(4C)}_{2112}(k_1,k_2,k_3) \nonumber \\
&&- \omega^2(k_1+k_2+k_3)G^{(4C)}_{1111}(k_1,k_2,k_3)
- \lambda G^{(2C)}_{11}(k_1)G^{(2C)}_{11}(k_2)G^{(2C)}_{11}(k_3) \nonumber \\
&&- \frac{\lambda}{2}\biggl[\int [dk^{\prime}]\biggl(
G^{(2C)}_{11}(k^{\prime})G^{(4C)}_{1111}(k_1,k_2,k_3)
+ G^{(2C)}_{11}(k_1)G^{(4C)}_{1111}(k^{\prime},k_2,k_3) \nonumber \\
&&+ G^{(2C)}_{11}(k_2)G^{(4C)}_{1111}(k^{\prime},k_1,k_3) +
G^{(2C)}_{11}(k_3)G^{(4C)}_{1111}(k^{\prime},k_1,k_2)
\biggr)\biggr], \\
 \dot{G}^{(4C)}_{2211}(k_1,k_2,k_3) &=&
G^{(4C)}_{2221}(k_1,k_2,k_3) +
G^{(4C)}_{2212}(k_1,k_2,k_3)-\omega^2(k_1+k_2+k_3)G^{(4C)}_{1211}(k_1,k_2,k_3)
- \omega^2(k_1)G^{(4C)}_{2111}(k_1,k_2,k_3) \nonumber \\
&&- \lambda \biggl(G^{(2C)}_{12}(k_1)G^{(2C)}_{11}(k_2)G^{(2C)}_{11}(k_3)
+ G^{(2C)}_{21}(k_1)G^{(2C)}_{11}(k_2)G^{(2C)}_{11}(k_3)\biggr) \nonumber \\
&&- \frac{\lambda}{2}\biggl[\int [dk^{\prime}]\biggl(
G^{(2C)}_{11}(k^{\prime})G^{(4C)}_{1211}(k_1,k_2,k_3) +
G^{(2C)}_{12}(k_1)G^{(4C)}_{1111}(k^{\prime},k_2,k_3)  \nonumber \\ &&+
G^{(2C)}_{11}(k_2)G^{(4C)}_{1121}(k^{\prime},k_1,k_3) +
G^{(2C)}_{11}(k_3)G^{(4C)}_{1121}(k^{\prime},k_1,k_2) \biggl) \nonumber \\ &&+
\int [dk^{\prime}]\biggl(G^{(2C)}_{11}(k^{\prime})
G^{(4C)}_{2111}(k_1,k_2,k_3)
+ G^{(2C)}_{21}(k_1)G^{(4C)}_{1111}(k^{\prime},k_2,k_3) \nonumber \\
&&+ G^{(2C)}_{11}(k_2)G^{(4C)}_{2111}(k^{\prime},k_1,k_3)  +
G^{(2C)}_{11}(k_3)G^{(4C)}_{2111}(k^{\prime},k_1,k_2)\biggr)\biggr],
\end{eqnarray}
\begin{eqnarray} \dot{G}^{(4C)}_{2221}(k_1,k_2,k_3) &=&
G^{(4C)}_{2222}(k_1,k_2,k_3)
- \omega^2(k_1+k_2+k_3)G^{(4C)}_{1221}(k_1,k_2,k_3) \nonumber \\
&&- \omega^2(k_1)G^{(4C)}_{2121}(k_1,k_2,k_3)
- \omega^2(k_2)G^{(4C)}_{2211}(k_1,k_2,k_3) \nonumber \\
&&- \lambda \biggl(G^{(2C)}_{12}(k_1)G^{(2C)}_{12}(k_2)G^{(2C)}_{11}(k_3)
+ G^{(2C)}_{21}(k_1)G^{(2C)}_{12}(k_2)G^{(2C)}_{11}(k_3) \nonumber \\
&&+ G^{(2C)}_{21}(k_1)G^{(2C)}_{21}(k_2)G^{(2C)}_{11}(k_3) \biggr) \nonumber \\
&&- \frac{\lambda}{2}\biggl[\int [dk^{\prime}]\biggl(
G^{(2C)}_{11}(k^{\prime})G^{(4C)}_{1221}(k_1,k_2,k_3)
+ G^{(2C)}_{12}(k_1)G^{(4C)}_{1121}(k^{\prime},k_2,k_3) \nonumber \\
&&+ G^{(2C)}_{12}(k_2)G^{(4C)}_{1121}(k^{\prime},k_1,k_3)
+ G^{(2C)}_{11}(k_3)G^{(4C)}_{1122}(k^{\prime},k_1,k_2)\biggr)  \nonumber \\
&&+ \int [dk^{\prime}]\biggl(
G^{(2C)}_{11}(k^{\prime})G^{(4C)}_{2121}(k_1,k_2,k_3)
+ G^{(2C)}_{21}(k_1)G^{(4C)}_{1121}(k^{\prime},k_2,k_3) \nonumber \\
&&+ G^{(2C)}_{12}(k_2)G^{(4C)}_{2111}(k^{\prime},k_1,k_3)
+ G^{(2C)}_{11}(k_3)G^{(4C)}_{2112}(k^{\prime},k_1,k_2)\biggr) \nonumber \\
&&+ \int [dk^{\prime}]\biggl(
G^{(2C)}_{11}(k^{\prime})G^{(4C)}_{2211}(k_1,k_2,k_3)
+ G^{(2C)}_{21}(k_1)G^{(4C)}_{2111}(k^{\prime},k_2,k_3) \nonumber \\
&&+ G^{(2C)}_{21}(k_2)G^{(4C)}_{2111}(k^{\prime},k_1,k_3) +
G^{(2C)}_{11}(k_3)G^{(4C)}_{2211}(k^{\prime},k_1,k_2)\biggr)\biggr],
\end{eqnarray}
\begin{eqnarray}
\dot{G}^{(4C)}_{2222}(k_1,k_2,k_3) &=& - \biggr(
\omega^2(k_1+k_2+k_3)G^{(4C)}_{1222}(k_1,k_2,k_3) +
\omega^2(k_1)G^{(4C)}_{2122}(k_1,k_2,k_3) \nonumber \\ &&+
\omega^2(k_2)G^{(4C)}_{2212}(k_1,k_2,k_3) +
\omega^2(k_3)G^{(4C)}_{2221}(k_1,k_2,k_3) \biggr) \nonumber \\
&&- \lambda \biggr( G^{(2C)}_{12}(k_1)G^{(2C)}_{12}(k_2)G^{(2C)}_{12}(k_3)
+ G^{(2C)}_{21}(k_1)G^{(2C)}_{12}(k_2)G^{(2C)}_{12}(k_3) \nonumber \\
&&+ G^{(2C)}_{21}(k_1)G^{(2C)}_{21}(k_2)G^{(2C)}_{12}(k_3)
+ G^{(2C)}_{21}(k_1)G^{(2C)}_{21}(k_2)G^{(2C)}_{21}(k_3) \biggr) \nonumber \\
&&- \frac{\lambda}{2}\biggr[\int [dk^{\prime}]\biggl(
G^{(2C)}_{11}(k^{\prime})G^{(4C)}_{1222}(k_1,k_2,k_3)
+ G^{(2C)}_{12}(k_1)G^{(4C)}_{1122}(k^{\prime},k_2,k_3) \nonumber \\
&&+ G^{(2C)}_{12}(k_2)G^{(4C)}_{1122}(k^{\prime},k_1,k_3)
+ G^{(2C)}_{12}(k_3)G^{(4C)}_{1122}(k^{\prime},k_1,k_2) \biggr) \nonumber \\
&&+ \int [dk^{\prime}]\biggl(
G^{(2C)}_{11}(k^{\prime})G^{(4C)}_{2122}(k_1,k_2,k_3)
+ G^{(2C)}_{21}(k_1)G^{(4C)}_{1122}(k^{\prime},k_2,k_3)  \nonumber \\
&&+ G^{(2C)}_{12}(k_2)G^{(4C)}_{2112}(k^{\prime},k_1,k_3)
+ G^{(2C)}_{12}(k_3)G^{(4C)}_{2112}(k^{\prime},k_1,k_2) \biggr) \nonumber \\
&&+ \int [dk^{\prime}]\biggl(
G^{(2C)}_{11}(k^{\prime})G^{(4C)}_{2212}(k_1,k_2,k_3)
+ G^{(2C)}_{21}(k_1)G^{(4C)}_{2112}(k^{\prime},k_2,k_3) \nonumber \\
&&+ G^{(2C)}_{21}(k_2)G^{(4C)}_{2112}(k^{\prime},k_1,k_3)
+ G^{(2C)}_{12}(k_3)G^{(4C)}_{2211}(k^{\prime},k_1,k_2) \biggr) \nonumber \\
&&+ \int [dk^{\prime}]\biggl(
G^{(2C)}_{11}(k^{\prime})G^{(4C)}_{2221}(k_1,k_2,k_3)
+ G^{(2C)}_{21}(k_1)G^{(4C)}_{2211}(k^{\prime},k_2,k_3) \nonumber \\
&&+ G^{(2C)}_{21}(k_2)G^{(4C)}_{2211}(k^{\prime},k_1,k_3) +
G^{(2C)}_{21}(k_3)G^{(4C)}_{2211}(k^{\prime},k_1,k_2)
\biggr)\biggr].
\end{eqnarray}

This set of equations completely determine the evolution of the
correlation functions beyond the leading order. Numerical integration
of this set of equations with nonequilibrium initial conditions
and tracking of the subsequent dynamical evolution of the
connected, n-point functions would make it possible to ascertain
whether the proposed truncation scheme is good enough to ensure
late-time thermalization of the system. The inclusion of the
4-point correlators in the hierarchy amounts to considering terms
of ${\cal O}(\hbar^3)$ in the effective action. Since each
connected n-point function $ G^{(nC)} \sim {\cal O}(\hbar^{n-1})$,
inclusion of higher order connected n-point correlators in the
hierarchy of evolution equations allows for a systematic way of
incorporating quantum effects. The issue of truncation error
arising from truncating the hierarchy at the 4-point level needs
to be addressed. This issue has been briefly discussed in Ref.
\cite{yaffe} where it was argued that truncation error would build
up with time and eventually invalidate the working assumption of
the formal hierarchy of the connected correlators given above. For
the quantum mechanical anharmonic oscillator and the O(N) vector
model, the decoherence time-scale for breakdown of the hierarchy
was qualitatively argued to scale with $\hbar^{-1/2}$ and
$\sqrt{N}$ respectively.  However, their analysis was based on a
simple $(0+1)$ dimensional quantum mechanical model and
quantitative treatment of this aspect for a system with infinite
degrees of freedom (field theory) will certainly be more
complicated. Although, inclusion of higher order n-point functions
in the hierarchy does not eliminate the instabilities, it does
postopone the onset of these instabilities. This delay in the 
appearance of intsabilities would make it possibile to gain valuable
insights into the non-equilibrium dynamics and approach to thermalization, 
in a temporal domain  which lies beyond the domain of validity of 
the Hartree approximation. A possible way to avoid the instabilities 
arising due to the truncation of the hierarchy might involve adapting the
Direct Interaction Approximation(DIA) developed by Kraichnan \cite{kra} 
in the context of turbulent fluid dynamics. We are currently exploring 
this possibility.

\section{Comparison with Alternative Methods}

 Several approaches have been developed to understand the dynamics of fields
in nonequilibrium field theory. It would be useful to make a
comparative study of the different approaches, specifying the
relationship between the different methods and the relative merits
and demerits of each approach. In this section, we will address
this issue by comparing our approach with other methods which have
been used to obtain evolution equations for the {\it equal-time}
correlation functions. We will compare our canonical approach with
two specific approaches \cite{yaffe,cw1} that have been recently
discussed in the literature.

Wetterich's method \cite{cw1} is based on deriving an evolution
equation for the partition function (or the generating functional
for 1PI graphs). The time evolution of the n-point correlation
functions (or n-point vertex functions) is completely determined by
the time evolution of the generating functional and the
microscopic dynamical equations of motion. We will show below that
the evolution equations for the equal-time correlation functions
obtained using Wetterich's method match exactly with those derived
in this paper using a canonical approach.

For a $\phi^4$-theory without spontaneous symmetry breaking, the
evolution equation for the generating functional can be cast in
the form \cite{cw1}
\begin{equation}
\partial_tZ[j(x),h(x);t] = ({\cal L}_{cl} + {\cal L}_{qm})Z[j(x),h(x);t]
\end{equation}
where ${\cal L}_{cl}$ and ${\cal L}_{qm}$ are the classical and
quantum parts of the Liouvillian operator respectively and are
given by
\begin{eqnarray}
{\cal L}_{cl} &=& \int d x \biggl[ j(x)\frac{\delta}{\delta h(x)}
+ h(x)\biggl\{\nabla^2 \frac{\delta}{\delta j(x)} - \biggl(
m^2\frac{\delta}{\delta j(x)} +
\frac{\lambda}{6}\frac{\delta^3}{\delta j(x)^3} \biggr) \biggr\}
\biggr] \nonumber \\
{\cal L}_{qm} &=& \frac{\lambda}{4!} \int d x
\biggl[h^3(x)\frac{\delta}{\delta j(x)}\biggr].
\end{eqnarray}
Here $j(x)$ and $h(x)$ are source terms for the field and
conjugate momenta, respectively, and $Z[j(x),h(x);t]$ is the
generating functional for n-point functions, $Z = {\rm
Tr}[\exp(\int d x(j(x)\phi(x)+h(x)\pi(x)))\rho]$, where $\rho$ is
the density matrix.

It is fairly straightforward to obtain the corresponding evolution
equations for the generating functional for connected graphs
$W[j(x),h(x);t]$ and the generating functional for 1PI graphs
$\Gamma[\varphi_c(x),\pi_c(x);t]$. ($\varphi_c(x)$ and $\pi_c(x)$
are vacuum expectation values of the field and conjugate
momentum in the presence of
sources $j(x)$ and $h(x)$, respectively). $Z[j(x),h(x);t],
W[j(x),h(x);t]$ and $\Gamma[\varphi_c(x),\pi_c(x);t]$ are related
by the equation $W[j(x),h(x);t] = \ln Z[j(x),h(x);t]$ and the
Legendre transform $\Gamma[\varphi_c(x),\pi_c(x);t] = - \ln
Z[j(x),h(x);t] + \int d x [j(x)\varphi_c(x) + h(x)\pi_c(x)]$. The
evolution equation for $W[j(x),h(x);t]$ which completely
determines the evolution of the connected n-point functions then
becomes
\begin{eqnarray}
\partial_tW[j(x),h(x);t] &=& \int dx \biggl[j(x)\frac{\delta W}{\delta h(x)}
+ h(x)(\nabla^2-m^2)\frac{\delta W}{\delta j(x)} -
\frac{\lambda}{6}h(x)\biggl\{\frac{\delta^3W}{\delta j(x)^3} +
3\frac{\delta W}{\delta j(x)}\frac{\delta^2W}{\delta j(x)^2}
+ \biggl(\frac{\delta W}{\delta j(x)}\biggr)^3 \biggr\} \biggr] \nonumber \\
&&+  \frac{\lambda}{4!}\int dx \biggl[h^3(x)\frac{\delta
W}{\delta j(x)}\biggr]. \label{eq cww}
\end{eqnarray}
For a symmetric theory $\varphi_c(x)|_{j=0} \equiv (\frac{\delta
W}{\delta j(x)})|_{j=0} = 0$ and $\pi_c(x)|_{h=0} \equiv
(\frac{\delta W}{\delta h(x)})|_{h=0} = 0$. Using Eq. (\ref{eq
cww}) we find that
\begin{eqnarray}
\dot{g}_{11}^{(2C)}({x_1},{x_2}) \equiv \frac{\delta^2}{\delta
j(x_1)\delta j(x_2)}\partial_tW[j,h;t]|_{j=h=0} &=&
g_{12}^{(2C)}({x_1},{x_2}) + g_{21}^{(2C)}({x_1},{x_2}),
\nonumber \\
\dot{g}_{21}^{(2C)}({x_1},{x_2}) \equiv \frac{\delta^2}{\delta
h(x_1)\delta j(x_2)}\partial_tW[j,h;t]|_{j=h=0} &=&
g_{22}^{(2C)}({x_1},{x_2}) +
\nabla_{1}^2g_{12}^{(2C)}({x_1},{x_2}) -
m^2g_{11}^{(2C)}({x_1},{x_2}) \nonumber \\ &&-
\frac{\lambda}{2}g_{11}^{(2C)}({x_1},{x_2})g_{11}^{(2C)}({x_1},{x_1})
- \frac{\lambda}{6}g_{1111}^{(4C)}({x_1},{x_1},{x_1},{x_2}),
\nonumber\\
\dot{g}_{22}^{(2C)}({x_1},{x_2}) \equiv \frac{\delta^2}{\delta
h(x_1)\delta h(x_2)}\partial_tW[j,h;t]|_{j=h=0} &=& (\nabla_{1}^2
- m^2)g_{12}^{(2C)}({x_1},{x_2})
+ (\nabla_{2}^2 - m^2)g_{21}^{(2C)}({x_1},{x_2}) \nonumber \\
&&- \frac{\lambda}{2} \biggl(
g_{11}^{(2C)}({x_1},{x_1})g_{12}^{(2C)}({x_1},{x_2})
+g_{11}^{(2C)}({x_2},{x_2})g_{21}^{(2C)}({x_1},{x_2})
\biggr) \nonumber \\
&&- \frac{\lambda}{6} \biggl(
g_{1112}^{(4C)}({x_1},{x_1},{x_1},{x_2}) +
g_{2111}^{(4C)}({x_1},{x_2},{x_2},{x_2}) \biggr).
\end{eqnarray}
The above set of equations for the connected 2-point function,
derived using Eq. (\ref{eq cww}) is identical to Eq. (A1)
which was derived using canonical methods and the cluster
expansion for ordinary n-point functions. Similarly, one can show
that the evolution for the connected 4-point function obtained
using the above method is the same as the ones obtained in
Appendix A.

More recently, Ryzhov and Yaffe (RY) developed another method for
obtaining the nonequilibrium, coupled set of evolution equations
for the n-point correlators \cite{yaffe}. Their method is based on
an expansion of the coherent state expectation value of the
products of operators in terms of subtracted n-point functions and
the assumption that the initial state is some coherent state. The
central idea behind the RY method is the expansion of the
Hamiltonian operator in terms of powers of the generators of
the underlying coherence group, which is the Heisenberg group for
the $\phi^4$-theory. It consists of the operators
{$e^{(1)},e^{(2)},e^{(3)}$} $\equiv$ {$\hat{\Phi}_{1}(k) , 
\hat{\Phi}_{2}(k) , \hat{1} $} satisfying the commutation relations 
$[e^{(1)},e^{(2)}] = if^{3}_{12}e^{(3)}\delta(k - k^{\prime})$ where
$f^{3}_{12}=-f^{3}_{21} = 1$ are the only non-vanishing structure
constants. By taking the {\it coherent-state} expectation value of
the Heisenberg equations of motion for appropriate products of
equal-time coherence group generators, it is possible to obtain
the evolution equations for the subtracted n-point functions (or
connected functions) in the same manner as described in Sec. V. In
Appendix B, the evolution equations for the two-point function are
derived in the Hartree approximation, using the RY method and are
shown to be equivalent to those obtained using the LvN approach.

\section{Conclusion}

In this paper we have used the LvN formalism, a canonical method,
to study the nonequilibrium evolution for the equal-time, connected 
n-point functions for a symmetric $\phi^4$-field theory. The usefulness 
and simplicity of the LvN method in obtaining perturbative, non-Gaussian 
corrections to the n-point correlation functions were first 
illustrated using the quantum-mechanical anharmonic oscillator model.  
The formalism was then applied to a $\phi^4$-field theory and used to obtain
the evolution equations for the connected 2-point function in the
Hartree approximation, for nonequilibrium evolution as well as for
the thermal equilibrium case. We also used the LvN formalism to 
go beyond the Gaussian approximation and obtain expressions for the 
2-point functions correct to ${\cal O}(\lambda^2)$ after calculating 
the improved vacuum state. Expressions for 4-point and higher 
n-point functions can be similarly obtained after some straightforward
but tedious algebra. The corrections to the Gaussian approximation 
were found to appear first at ${\cal O}(\lambda^2)$. The nonequilibrium
evolution equations beyond the leading order approximation were
then obtained by taking the vacuum expectation value of the
Heisenberg equations of motion for appropriate products of field
operators. This provides an alternative and non-perturbative
approach to  investigate systematically the nonequilibrium
evolution of quantum fields and yielded an infinite hierarchy of
coupled equations for the connected n-point correlators which were
truncated at the 4-point level. This involved ignoring the
contribution of connected n-point functions ($n \ge 6$) in the
evolution and resulted in a closed set of equations involving
connected 2-point and 4-point functions. Since we restricted our
investigation to a symmetric $\phi^4$-theory with vanishing vacuum
expectation value, all connected odd-n-point functions vanish and
have no effect on the evolution. We have also established a
connection between the canonical approach used in this paper and
other methods \cite{yaffe,cw1} developed in the literature for
deriving nonequilibrium evolution equations for equal-time
correlation functions. It would be straightforward to generalize
this technique for studying the nonequilibrium evolution of 
spontaneously broken field theories. Symmetry breaking would
result in the generation of linear and cubic terms in the
potential, as a consequence of which, even connected odd-n-point
functions would contribute to the dynamical evolution.

The inadequacy of the Hartree approximation in studying the
approach to thermalization has been extensively discussed in the
literature. The Hartree approximation neglects scattering effects
and therefore cannot account for long-time thermalization of the
system. An elegant interpretation of this aspect has been put
forward by Wetterich \cite{cw1} and is based on the realization
that the Hartree solution corresponds to a fixed point of the
theory (i.e. configurations for which $\partial_tZ=0$). The
presence of an infinite number of conserved quantities (fixed
points) prevents the system from escaping from these
nonequilibrium fixed points and approaching the thermal
equilibrium fixed point unless the initial values of all these
conserved quantities coincide exactly with the ones corresponding
to a thermal distribution. To take into account the effect of
scattering which would ultimately lead to thermalization of the
system, requires going beyond the Hartree approximation. An approach 
based on the loop expansion of the 2PI effective action \cite{berges} 
in powers of $\hbar$ (or $1/N$ for an O(N) symmetric field theory) holds 
more promise in this context. In this paper we have discussed two different 
canonical approaches (perturbative and non-perturbative) of going beyond 
the Hartree approximation. The perturbative LvN approach developed in 
this paper provides an elegant method for obtaining corrections 
to the n-point functions in terms of powers of the coupling constant
$\lambda$. The effect of scattering which first appears at 
${\cal O}(\lambda^2)$ is manifest through the first and second order
coefficients whose dynamical equations were derived. This method provides a  
straightforward method for obtaining non-Gaussian corrections to the 
domain size in systems undergoing a quenched second order phase 
transition. By making use of the exponentially growing solution of the 
soft modes of the theory, it is possible to obtain expressions for the 
first and second order coefficients, which in turn lead to expressions 
of the 2-point functions to various orders in $\lambda$. The Fourier
transform of the 2-point functions would then yield the domain size 
which includes corrections due to non-Gaussian effects \cite{kim-khanna2}. 
Moreover, the LvN method provides an analytical, albeit perturbative, method 
for studying the role of interactions and effect of scattering on thermalization 
of the system. This is possible by obtaining perturbative corrections to the 
2-point functions to various powers of $\lambda$ and comparing the 
resulting expression with the thermal 2-point correlation functions. 
We have also discussed in detail an alternative, non-perturbative 
canonical approach based on the Heisenberg formalism to study the nonequilibrium
evolution of the $\phi^4$ field theory. This non-perturbative Heisenberg formalism 
requires incorporating the effect of connected 4-point function in the
nonequilibrium evolution equations. The efficacy of truncating the
hierarchy at the 4-point level in ensuring thermalization needs to
be addressed. Recent work \cite{aarts} for classical field
theories has been inconclusive because of the prohibitively long
time required for equilibration and also due to the fact that the
numerical evolution becomes unstable long before the system
thermalizes. We believe that the canonical approaches developed in this paper 
provide an useful alternative method for studying nonequilibrium dynamics 
in field theory.

\acknowledgements

S.S. would like to thank L. Yaffe for helpful discussions. S.P.K.
would like to thank D.N. Page for useful discussions and also
appreciate the warm hospitality of the Theoretical Physics
Institute, University of Alberta. The work of S.S. and F.C.K. was
supported in part by the Natural Sciences and Engineering Research
Council (NSERC), Canada. The work of S.P.K. was supported in part
by the  Korea Research Foundation under Grant No. KRF-2002-041-C00053 
and also by the Korea Science and Engineering Foundation under 
Grant No. 1999-2-112-003-5.

\appendix
\section{}

In this appendix we write down the set of evolution equations for
the connected n-point functions in configuration space by making
use of the cluster expansion for the 4-point and 6-point
functions. The cluster expansion of n-point functions can be
formally derived from the generating functional of the interacting
field theory \cite{peter}. The difference between the ordinary and
connected n-point correlators first appears at the 4-point level.
To simplify the notation, we represent the $\phi$ and $\pi$ field
operators by the numbers 1 and 2 respectively. So the correlation
functions $g^{(2C)}_{\phi\phi}(x_1,x_2)$, $g^{(2C)}_{\pi\phi}(x_1,
x_2)$ etc. are symbolically represented as $g^{(2C)}_{11}(1,2)$,
$g^{(2C)}_{21}(1,2)$, respectively, where the numbers in the
brackets refer to the subscript indices of the spatial
coordinates.

The equations for the 2-point functions in configuration space are
easily obtained by using Eq. (\ref{eveq1}) and the appropriate
cluster expansion of the 4-point functions. They are
\begin{eqnarray}
\dot{g}^{(2C)}_{11}(1,2) &=& g^{(2C)}_{12}(1,2) +
g^{(2C)}_{12}(1,2), \nonumber \\ \nonumber \\
\dot{g}^{(2C)}_{21}(1,2) &=& g^{(2C)}_{22}(1,2) + (\nabla_{1}^2 -
m^2)g^{(2C)}_{12}(1,1) - \frac{\lambda \ 
}{2}g^{(2C)}_{11}(1,1)g^{(2C)}_{11}(1,2) - \frac{\lambda \ 
}{6}g^{(4C)}_{1111}(1,1,1,2), \nonumber \\ \nonumber
\\ \dot{g}^{(2C)}_{22}(1,2) &=& (\nabla_{1}^2 -
m^2)g^{(2C)}_{12}(1,2) + (\nabla_2^2 - m^2)g^{(2C)}_{21}(1,2) -
\frac{\lambda \ }{2}\biggl(g^{(2C)}_{11}(1,1)g^{(2C)}_{12}(1,2)
+ g^{(2C)}_{11}(2,2)g^{(2C)}_{21}(1,2) \biggr) \nonumber \\ &&-
\frac{\lambda \ }{6}\biggl(g^{(4C)}_{1112}(1,1,1,2) +
g^{(4C)}_{2111}(1,2,2,2) \biggr),
\end{eqnarray}
where the last two equations in the above set have been obtained
by making use of the following cluster expansion of the 4-point
functions
\begin{eqnarray}
g^{(4)}_{1111}(1,1,1,2) &=& g^{(4C)}_{1111}(1,1,1,2) +
3 g^{(2C)}_{11}(1,1)g^{(2C)}_{11}(1,2), \nonumber \\
g^{(4)}_{1112}(1,1,1,2) &=& g^{(4C)}_{1112}(1,1,1,2) +
3 g^{(2C)}_{11}(1,1)g^{(2C)}_{12}(1,2), \nonumber \\
g^{(4)}_{2111}(1,2,2,2) &=& g^{(4C)}_{2111}(1,1,1,2) + 3
g^{(2C)}_{11}(2,2)g^{(2C)}_{21}(1,2).
\end{eqnarray}
We have used lower case letters to represent the correlation
functions in configuration space and upper case letters to
represent their Fourier transformed counterpart in momentum space.
The n-point connected correlation functions appearing above and in
the subsequent discussion are all normal ordered vacuum
expectation values of products of operators.

The evolution equations for the connected 4-point functions in
configuration space can be similarly derived by making use of the
appropriate cluster expansion of the 6-point functions. The five
independent equations for the 4-point function are
\begin{equation}
\dot{g}^{(4C)}_{1111}(1,2,3,4) = g^{(4C)}_{2111}(1,2,3,4) +
g^{(4C)}_{1211}(1,2,3,4) + g^{(4C)}_{1121}(1,2,3,4) +
g^{(4C)}_{1112}(1,2,3,4),
\end{equation}
\begin{eqnarray}
\dot{g}^{(4C)}_{2111}(1,2,3,4) &=& g^{(4C)}_{2211}(1,2,3,4) +
g^{(4C)}_{2121}(1,2,3,4) + g^{(4C)}_{2112}(1,2,3,4) + (\nabla_1^2
- m^2)g^{(4C)}_{1111}(1,2,3,4) \nonumber \\ &&- \lambda \ 
g^{(2C)}_{11}(1,2)g^{(2C)}_{11}(1,3)g^{(2C)}_{11}(1,4) -
\frac{\lambda \ }{2}
\biggl(g^{(2C)}_{11}(1,1)g^{(4C)}_{1111}(1,2,3,4) +
g^{(2C)}_{11}(1,2)g^{(4C)}_{1111}(1,1,3,4) \nonumber \\ &&+
g^{(2C)}_{11}(1,3)g^{(4C)}_{1111}(1,1,2,4) +
g^{(2C)}_{11}(1,4)g^{(4C)}_{1111}(1,1,2,3) \biggr),
\end{eqnarray}
\begin{eqnarray}
\dot{g}^{(4C)}_{2211}(1,2,3,4) &=& g^{(4C)}_{2221}(1,2,3,4) +
g^{(4C)}_{2212}(1,2,3,4) + (\nabla_1^2 -
m^2)g^{(4C)}_{1211}(1,2,3,4) + (\nabla_2^2 -
m^2)g^{(4C)}_{2111}(1,2,3,4) \nonumber \\ &&- \lambda \ 
\biggl(g^{(2C)}_{12}(1,2)g^{(2C)}_{11}(1,3)g^{(2C)}_{11}(1,4) +
g^{(2C)}_{21}(1,2)g^{(2C)}_{11}(2,3)g^{(2C)}_{11}(2,4) \biggr)
\nonumber \\ &&- \frac{\lambda \ 
}{2}\biggl(g^{(2C)}_{11}(1,1)g^{(4C)}_{1211}(1,2,3,4) +
g^{(2C)}_{12}(1,2)g^{(4C)}_{1111}(1,1,3,4) \nonumber \\ &&+
g^{(2C)}_{11}(1,3)g^{(4C)}_{1121}(1,1,2,4) +
g^{(2C)}_{11}(1,4)g^{(4C)}_{1121}(1,1,2,3) \biggr) \nonumber \\
&&- \frac{\lambda \ 
}{2}\biggl(g^{(2C)}_{21}(1,2)g^{(4C)}_{1111}(2,2,3,4) +
g^{(2C)}_{11}(2,2)g^{(4C)}_{2111}(1,2,3,4) \nonumber \\ &&+
g^{(2C)}_{11}(2,3)g^{(4C)}_{2111}(1,2,2,4) +
g^{(2C)}_{11}(2,4)g^{(4C)}_{2111}(1,2,2,3) \biggr).
\end{eqnarray}

Since the correlation functions are all normal ordered, the above
equation is symmetric under the interchange of indices 
$x_1 \leftrightarrow x_2$ and $x_3 \leftrightarrow x_4$
\begin{eqnarray}
\dot{g}^{(4C)}_{2221}(1,2,3,4) &=& g^{(4C)}_{2222}(1,2,3,4) +
(\nabla_1^2 - m^2)g^{(4C)}_{1221}(1,2,3,4) + (\nabla_2^2 -
m^2)g^{(4C)}_{2121}(1,2,3,4) \nonumber \\ &&+ (\nabla_3^2 -
m^2)g^{(4C)}_{2211}(1,2,3,4) - \lambda \ 
\biggl(g^{(2C)}_{12}(1,2)g^{(2C)}_{12}(1,3)g^{(2C)}_{11}(1,4)
\nonumber \\ &&+
g^{(2C)}_{21}(1,2)g^{(2C)}_{12}(2,3)g^{(2C)}_{11}(2,4) +
g^{(2C)}_{21}(1,3)g^{(2C)}_{21}(2,3)g^{(2C)}_{11}(3,4) \biggr)
\nonumber \\ &&- \frac{\lambda \ 
}{2}\biggl(g^{(2C)}_{11}(1,1)g^{(4C)}_{1221}(1,2,3,4) +
g^{(2C)}_{12}(1,2)g^{(4C)}_{1111}(1,1,3,4) \nonumber \\ &&+
g^{(2C)}_{12}(1,3)g^{(4C)}_{1121}(1,1,2,4) +
g^{(2C)}_{11}(1,4)g^{(4C)}_{1122}(1,1,2,3) \biggr)  \nonumber \\
&&- \frac{\lambda \ 
}{2}\biggl(g^{(2C)}_{11}(1,2)g^{(4C)}_{2121}(1,2,3,4) +
g^{(2C)}_{21}(1,2)g^{(4C)}_{1121}(2,2,3,4) \nonumber \\ &&+
g^{(2C)}_{12}(1,3)g^{(4C)}_{2111}(1,2,2,4) +
g^{(2C)}_{11}(2,4)g^{(4C)}_{2112}(1,2,2,3) \biggr)  \nonumber \\
&&- \frac{\lambda \ 
}{2}\biggl(g^{(2C)}_{11}(3,3)g^{(4C)}_{2211}(1,2,3,4) +
g^{(2C)}_{21}(1,3)g^{(4C)}_{2111}(2,3,3,4) \nonumber \\ &&+
g^{(2C)}_{21}(1,3)g^{(4C)}_{2111}(1,3,3,4) +
g^{(2C)}_{11}(3,4)g^{(4C)}_{2211}(1,2,3,3) \biggr).
\end{eqnarray}
The above equation is symmetric under the exchange of coordinate
indices $x_1 \leftrightarrow x_2 \leftrightarrow x_3$.
\begin{eqnarray}
\dot{g}^{(4C)}_{2222}(1,2,3,4) &=& (\nabla_1^2 -
m^2)g^{(4C)}_{1222}(1,2,3,4) + (\nabla_2^2 -
m^2)g^{(4C)}_{2122}(1,2,3,4) \nonumber \\ &&+ (\nabla_3^2 -
m^2)g^{(4C)}_{2212}(1,2,3,4) + (\nabla_4^2 -
m^2)g^{(4C)}_{2221}(1,2,3,4) \nonumber \\ &&- \lambda \ 
\biggl(g^{(2C)}_{12}(1,2)g^{(2C)}_{12}(1,3)g^{(2C)}_{12}(1,4) +
g^{(2C)}_{21}(1,2)g^{(2C)}_{12}(2,3)g^{(2C)}_{12}(2,4) \nonumber
\\ &&+ g^{(2C)}_{21}(1,3)g^{(2C)}_{21}(2,3)g^{(2C)}_{12}(3,4) +
g^{(2C)}_{21}(1,4)g^{(2C)}_{21}(2,4)g^{(2C)}_{21}(3,4) \biggr)
\nonumber \\ &&- \frac{\lambda \ 
}{2}\biggl(g^{(2C)}_{11}(1,1)g^{(4C)}_{1222}(1,2,3,4) +
g^{(2C)}_{12}(1,2)g^{(4C)}_{1122}(1,1,3,4) \nonumber \\ &&+
g^{(2C)}_{12}(1,3)g^{(4C)}_{1122}(1,1,2,4) +
g^{(2C)}_{12}(1,4)g^{(4C)}_{1122}(1,1,2,3) \biggr) \nonumber \\
&&- \frac{\lambda \ 
}{2}\biggl(g^{(2C)}_{21}(1,2)g^{(4C)}_{1122}(2,2,3,4) +
g^{(2C)}_{11}(2,2)g^{(4C)}_{2122}(1,2,3,4) \nonumber \\ &&+
g^{(2C)}_{12}(2,3)g^{(4C)}_{2112}(1,2,2,4) +
g^{(2C)}_{12}(2,4)g^{(4C)}_{2112}(1,2,2,3) \biggr) \nonumber \\
&&-\frac{\lambda \ 
}{2}\biggl(g^{(2C)}_{21}(1,3)g^{(4C)}_{2112}(2,3,3,4) +
g^{(2C)}_{21}(2,3)g^{(4C)}_{2112}(1,3,3,4) \nonumber \\ &&+
g^{(2C)}_{11}(3,3)g^{(4C)}_{2212}(1,2,3,4) +
g^{(2C)}_{12}(3,4)g^{(4C)}_{2211}(1,2,3,3) \biggr) \nonumber \\
&&- \frac{\lambda \ 
}{2}\biggl(g^{(2C)}_{21}(1,4)g^{(4C)}_{2211}(2,3,4,4) +
g^{(2C)}_{21}(2,4)g^{(4C)}_{2211}(1,3,4,4) \nonumber \\ &&+
g^{(2C)}_{21}(3,4)g^{(4C)}_{2211}(1,2,4,4) +
g^{(2C)}_{11}(4,4)g^{(4C)}_{2221}(1,2,3,4) \biggr).
\end{eqnarray}

\section{}

We now show the equivalence between the LvN method and the RY
method by obtaining the evolution equations for the Fourier
transform of the subtracted 2-point correlations using the RY
method in the Hartree approximation. The expectation value of the
Hamiltonian, (\ref{ham sum}), can be expressed in terms of the
Fourier transforms, $\Phi_{0i}(k, t), (i=1,2)$, of the coherent
state expectation value of the fluctuation field $\phi_{0}(x,t)
\equiv \langle \hat{\phi}_{f} \rangle_{\rm cs}$ and that of
conjugate momentum operator $\pi_{0}(x,t) \equiv \langle
\hat{\pi}_{f} \rangle_{\rm cs}$, respectively. In order to apply
the RY method, it is necessary to isolate from the expectation
value of the Hamiltonian, the parts which depend only on the
Fourier transforms $\Phi_{01}(k, t)=\Phi_{0}(k,t)$ and
$\Phi_{02}(k, t)=\Pi_{0}(k, t)$ of the coherent state expectation
value of the field and its conjugate momentum operator. Using the
relations $\langle \hat{\pi}_{f}^{2}(x)\rangle_{\rm cs} =
\pi_{0}^{2}(x,t) + g_{22}(0,t)$ and $\langle
\hat{\phi}_{f}^{2}(x)\rangle_{\rm cs} = \phi_{0}^{2}(x,t) +
g_{11}(0,t)$, and the Hartree factorization of the cubic and
quartic terms specified in Eq. (\ref{eq ha}), the coherent state
expectation value of the full Hamiltonian can be expressed as
\begin{equation}
\langle \hat{H}(t)\rangle_{\rm cs} = \int [d k]
\bar{H}(\Phi_{0}(k,t), \Pi_{0}(k,t)) + \cdots,
\end{equation}
where the dots correspond to terms which are independent of
$\Phi_{0i}(k,t), (i=1,2)$, and
\begin{equation}
\bar{H}(\Phi_{0}(k,t),\Pi_{0}(k,t)) = \frac{1}{2}\Pi_{0}^{2}(k,t)
+ \Bigl[ \omega_{k}^{2} + \frac{\lambda}{2}\phi_{cl}^2(t) +
\frac{\lambda}{2}g_{11}(0,t)\Bigr] \frac{\Phi_{0}^{2}(k,t)}{2}.
\label{ryh}
\end{equation}

By noting that the underlying coherence group in this case is the
Heisenberg group whose generators {$e^{(1)},e^{(2)},e^{(3)}$}
$\equiv$ {$\hat{\pi}_{k} , \hat{\phi}_{k} ,
\hat{1}$} satisfy the commutation relations
$[e^{(1)},e^{(2)}] = i f^{3}_{12}e^{(3)}\delta(k -  k')$ where
$f^{3}_{12} =-f^{3}_{21}=1$ are the only nonvanishing structure
constants, and by making use of the general form of the NLO
evolution equations for the 1-point and 2-point functions, we get
the following self-consistent set of equations to
${\cal{O}}((\lambda\ )^2)$
\begin{eqnarray}
\frac{d\Phi_{0}(k,t)}{dt} &=& \Pi_{0}(k,t), \nonumber
\\ \frac{d\Pi_{0}(k,t)}{dt} &=& - \Bigl[\omega_{k}^{2} +
\frac{\lambda}{2}\phi_{cl}^2(t) + \frac{\lambda}{2}g_{11}({0},t)
\Bigr]\Phi_{0}({k},t) + {\cal{O}}((\lambda \ )^2), \nonumber \\
\frac{dG_{11}({k},t)}{dt} &=& G_{12}({k},t) + G_{21}({k},t),
\nonumber\\ \frac{dG_{22}({k},t)}{dt} &=& - \Bigl[\omega_{k}^{2} +
\frac{\lambda}{2}\phi_{cl}^2(t) + \frac{\lambda}{2}g_{11}(
0,t)\Bigr](G_{12}(k,t) + G_{21}(k,t)) + {\cal{O}}((\lambda
\ )^2), \nonumber \\ \frac{dG_{12}(k,t)}{dt} &=& G_{22}(k,t) -
\Bigl[\omega_{k}^{2} + \frac{\lambda}{2}\phi_{cl}^2(t) +
\frac{\lambda}{2}g_{11}(0,t) \Bigr]G_{11}(k,t) +
{\cal{O}}((\lambda \ )^2), \label{eq ryhat}
\end{eqnarray}
where we have used $\bar{H}^{(11)} \equiv
\frac{1}{2!}\frac{\partial^2\bar{H}}{\partial \Phi_{0}^{2}(k,t)} =
\frac{1}{2}[\omega_{k}^{2} + \frac{\lambda}{2}\phi_{cl}^2(t) +
\frac{\lambda}{2}g_{11}(0,t)]$ and $\bar{H}^{(22)} \equiv
\frac{1}{2!}\frac{\partial^2\bar{H}}{\partial \Pi_{0}^{2}(k,t)} =
\frac{1}{2}$. The set of equations (\ref{eq ryhat}) are identical
to the equations (\ref{eq gcor}) obtained using the LvN formalism
which clearly establishes the equivalence of the LvN and RY
methods.  The coherent state can be considered as the vacuum state
of the theory with its expectation value providing the classical
background. For a symmetric theory, the coherent state expectation
value of the field and the conjugate momenta vanishes, and we
recover the set of equations for the 2-point functions (in the
Hartree limit) obtained in Sec. V.

\end{document}